\documentclass[referee,usenatbib]{mn2e}
\usepackage{epsfig}
\usepackage{subfigure}
\usepackage{color}
\newcommand{\Msun}{\>{\rm M_{\odot}}}
\newcommand{\msun}{\>{\rm M_{\odot}}}

\newcommand{\beq}{\begin{equation}}
\newcommand{\eeq}{\end{equation}}

\newcommand{\Obaryon}{{\Omega_{\rm B,0}}}

\newcommand{\msunh}{\>h^{-1}\rm M_\odot}

\def\Msun{\,\rm M_{\odot}}

\newcommand{\rvir}{r_{\rm vir}}
\newcommand{\vvir}{v_{\rm vir}}

\newcommand{\mcooldot}{\dot{m}_{\rm cool}}

\newdimen\hssize
\newdimen\hdsize 
\begin{document}
%%%%%%%%%%%%%%%%%%%%%%%%%%%%%%%%%%%%%%%%%%%%%%%%%%%%%%%%%%%%%%%%%%%%%%%%%%
\title[BIE-SAM]
      {A Bayesian approach to the semi-analytic model of galaxy 
       formation: methodology}
\author[]
       {
        Yu Lu$^{1,2}$\thanks{E-mail: luyu@stanford.edu},
		H.J. Mo$^{1}$, Martin D. Weinberg$^1$, Neal Katz$^1$
\\
        $^1$ Department of Astronomy, University of Massachusetts,
        Amherst, MA 01003-9305, USA
\\
		$^2$ Kavli Institute for Particle Astrophysics and Cosmology, Stanford, CA 94309, USA
        }

%%%%%%%%%%%%%%%%%%%%%%%%%%%%%%%%%%%%%%%%%%%%%%%%%%%%%%%%%%%%%%%%%%%%%%%%%%

\date{}

\pagerange{\pageref{firstpage}--\pageref{lastpage}}
\pubyear{2010}

\maketitle

%%%%%%%%%%%%%%%%%%%%%%%%%%%%%%%%%%%%%%%%%%%%%%%%%%%%%%%%%%%%%%%%%%%%%%%%%%
\begin{abstract}
  We believe that a wide range of physical processes conspire to shape
  the observed galaxy population but we remain unsure of their
  detailed interactions.  The semi-analytic model (SAM) of galaxy
  formation uses multi-dimensional parameterisations of the physical
  processes of galaxy formation and provides a tool to constrain these
  underlying physical interactions.  Because of the high
  dimensionality, the parametric problem of galaxy formation may be
  profitably tackled with a Bayesian-inference based approach, which
  allows one to constrain theory with data in a statistically rigorous
  way.
  In this paper we develop a SAM in the framework of
  Bayesian inference.  We show that, with a parallel implementation of
  an advanced Markov-Chain Monte-Carlo algorithm, it is now possible
  to rigorously sample the posterior distribution of the
  high-dimensional parameter space of typical SAMs.  As an example, we
  characterise galaxy formation in the current $\Lambda$CDM cosmology
  using the stellar mass function of galaxies as an observational
  constraint.  We find that the posterior probability distribution is
  both topologically complex and degenerate in some important model
  parameters, suggesting that thorough explorations of the parameter 
  space are needed to understand the models.  
  We also demonstrate that because of the model degeneracy,  
  adopting a narrow prior strongly restricts the model. Therefore, 
  the inferences based on SAMs are conditional to the model adopted. 
  Using synthetic data
  to mimic systematic errors in the stellar mass function, we
  demonstrate that an accurate observational error model is essential
  to meaningful inference.
\end{abstract}

%%%%%%%%%%%%%%%%%%%%%%%%%%%%%%%%%%%%%%%%%%%%%%%%%%%%%%%%%%%%%%%%%%%%%%%%%%
%%%%%%%%%%%%%%%%%%%%%%%%%%%%%%%%%%%%%%%%%%%%%%%%%%%%%%%%%%%%%%%%%%%%%%%%%%
\section{Introduction}

In our current paradigm of structure formation, the matter density of
the Universe is dominated by cold dark matter (hereafter CDM), and
galaxy formation is a two-stage process \citep[e.g.][]{White1978}.
First, small perturbations in the density field, originating from
quantum fluctuations in the early universe, grow and produce a
population of virialised dark matter halos. Second, the baryonic
matter associated with these halos accumulates at the halo centres owing
to cooling and cold flows, forming stars and galaxies. Because of the
hierarchical nature of structure formation in a CDM cosmogony, dark
matter halos merge. The halo mergers eventually lead to galaxy-galaxy 
mergers, resulting in the formation of elliptical galaxies.

The first stage of this process, the formation and virialisation of
dark matter halos, has been studied in great detail using the
(extended) Press-Schechter formalism \citep[e.g.][]{Press1974,
  Bond1991, Lacey1993}, spherical and ellipsoidal collapse
\citep[e.g.][]{Gunn1972, Fillmore1984, Bertschinger1985, Avila-Reese1998, Sheth2001,
  Lu2006} and numerical simulations \citep[e.g.][]{Efstathiou1985,
  Navarro1997, Bullock2001, Bullock2001a, Zhao2003a, Zhao2003,  
Springel2005a,Maccio2007,Zhao2009}. 
These studies have yielded the mass function, spatial
distribution, formation history, and internal structure of the CDM
halo population and serve as the backbone for any study of galaxy
formation.  The knowledge of the second stage of galaxy formation 
is far less well
established, mainly because the baryonic processes involved (cooling,
star-formation and feedback) are poorly understood.  Additional
physical processes whose importance is not fully understood include
dynamical friction, tidal stripping, black hole formation and
accretion, and adiabatic contraction.

Hydrodynamic simulations can now be used to study galaxy formation
and evolution in a full cosmological context \citep[e.g.][]{Katz1992,
  Navarro1993, Kerevs2005, Oppenheimer2006, Simha2009}.  However,
computational power is still a severe limitation at the present, and
one has to compromise between simulation resolution and box size.
Because of this, an alternative approach, the semi-analytical model of
galaxy formation, has been developed and widely adopted to study the
statistical properties of the galaxy population
\citep[e.g.][]{White1991, Kauffmann1993, Mo1998, Somerville1999,
  Avila-Reese2000, Cole2000, Firmani2000, Kang2005, Croton2006, Dutton2009}.  
In the semi-analytical
model (hereafter SAM), one adopts ``recipes'' to describe and
parametrise the underlying physical ingredients, such as star
formation and feedback. The free parameters in the models are then
tuned to reproduce certain observational data of the galaxy
population, such as stellar mass functions, colour-magnitude relations,
metallicity-stellar mass relations, Tully-Fisher relation, and
two-point statistics that describe the spatial distribution of
galaxies (e.g. the two-point correlation function, the pairwise peculiar
velocity dispersion, etc.).  However, the theory of galaxy formation
and evolution still faces several outstanding problems \citep[see][for
an up-to-date review]{Primack2009}. For example, it remains a
challenge to fit the faint-end slope of the galaxy luminosity
function \citep[e.g.][]{Benson2003, Mo2005}, and the models typically
predict disk rotation velocities that are too high, unless adiabatic
contraction and/or disk self-gravity are ignored
\citep[e.g.][]{Cole2000, Dutton2007}. In addition, the models have
problems matching the evolution of the galaxy mass function with
redshift \citep[e.g.][]{DeLucia2007, Somerville2008, Fontanot2009},
and typically overpredict the fraction of red satellite galaxies
\citep{Baldry2006, Weinmann2006, Kimm2009, Liu2010}. There are three
main reasons for these problems. First and foremost, current models
most likely miss some vital ingredients or the recipes used do not
properly implement the physical mechanism.  Second, sub-space
features and degeneracies in the model parameter space have been
either missed or not sufficiently explored \citep{Liu2010,Neistein2010}.  
Third, the difficulties may actually reflect inconsistencies in the data
themselves (so-called ``systematic'' errors). For example, it has been
pointed out that the observed evolution in the stellar mass function
is inconsistent with the observed cosmic star formation history
\citep{Fardal2007, Primack2008}.

To address these problems, one must quantitatively
characterise the model constraints implied by existing data sets as
well as explore a wider range of models.  The SAM approach provides a
promising avenue to tackle these problems owing to its flexibility in
implementation and its relatively fast speed in computation.  However,
significant changes in the methodology must be made to fully
utilise the potential of SAMs.  
The main shortcoming in current
SAM  implementations is that they are not probabilistically rigorous.  In many
published SAM applications, a subset of model parameters are held fixed
while other parameters are adjusted to match some observational
properties.  If the match is unsatisfactory, one further adjusts some
of the parameters or changes the model parametrisation until a
``good'' fit is achieved.  However, the goodness of fit is often
assessed ``by eye''; one overlays the model prediction, a luminosity
function for example, on the observed result to see if the prediction
is sufficiently close to the data.  Since the statistical
uncertainties in both the data and the model are not consistently
computed, confidence levels do not follow.  
Similarly, since the model
parameters are explored by hand, marginal probability can not be
computed.  As mentioned earlier, a number of physical processes in
galaxy formation are still poorly understood, and so the
parameterisations of these processes have to be made very general.
This leaves a large parameter space to be probed.  Given the high
dimensionality of the parameter space and the complex covariance
between parameters, it is almost impossible to find and delineate the
dominant mode by hand-tuning model parameters.  
%For example, if a good
%fit between model and data is not found, the significance of
%the parameter region relative to others remains unknown.  
Third, since
the model parameters might be strongly covariant, the effect of changing
one model parameter is conditional to the
values of other parameters that are kept fixed.  Therefore,
switching on and off a process in a fiducial model is unlikely to 
determine its importance to galaxy formation.  Indeed, to investigate the
influence of a specific recipe, one should allow the parameters to
range over their entire a priori plausible domain and marginalise 
over all the other parameters. 
Unfortunately, this kind of analysis has not been commonly adopted 
owing to the lack of suitable methods. 
Fourth, because many processes in galaxy formation are still
poorly understood, different SAMs may adopt different
parameterisations for the same process. While all these models can be
tuned to match a limited set of observational data 
and they are all considered as ``plausible'', whether 
one model is favoured by the data more than another needs to be 
assessed by the marginal likelihood, instead of by the 
goodness-of-fit of a single optimised parameter set. To 
compute the marginal likelihood or evidence, all the parameters 
should be allowed to vary in the domain specified by the priors, 
so that model selection can be made according to statistical 
evidence. Again, such an analysis is not included in the
current SAMs.

In summary, a variety of physical processes affecting galaxy formation
are not yet well understood while copious observational data exist to constrain
the models.  Thus, to derive meaningful constraints from
observations, we would like to know the probability of the various
model parameters and, indeed, entire model families given the data.
This leads us directly to Bayesian inference!  The semi-analytical
model provides a very powerful tool to translate the theory of galaxy
formation into a set of model parameters. The Bayesian approach will
then allow us to obtain the posterior distribution of the model
parameters for a given set of data and to assess how a particular
model is supported by the data. Moreover, given different model
families, Bayesian model comparison techniques such as Bayes Factors
and Reversible Jump techniques \citep{Green1995} allow one to 
determine the relative odds for each model to reproduce the observed data. 

Some attempts have been made recently in this direction.  For
instance, \citet{Bower2010} have explored the parameter space 
of the GALFORM model \citep{Bower2006} using a model emulator 
based on Latin hypercube sampling \citep{Mckay1979}, 
and identified a small fraction of the initial volume of 
the parameter space that is not ruled out by their using
the $K$- and $b_{\rm J}$-band luminosity functions of
galaxies in the local Universe as constraints. Using the same technique, 
\citet{Benson2010a} have performed an exhaustive search of model 
parameter space constrained by more observational data.
\citet{Kampakoglou2008} and \citet{Henriques2009} have
adopted the Markov-Chain Monte-Carlo (MCMC) technique to 
explore the ability of their adopted SAMs to accommodate 
multiple observational data sets.  

In this paper, we develop a scheme to incorporate SAMs into the
framework of Bayesian inference. 
To this end, we generalise the parameterisations for the model recipes 
so that our model can encompass the large uncertainties owing to our
limited knowledge of galaxy formation. 
We also show that, aided with advanced MCMC techniques and moderate
computational facilities, it is now possible to build a Bayesian
inference-based SAM to efficiently explore the high dimensional
parameter space involved and to establish the posterior distribution
of model parameters reliably.

The goal of the present paper is a description of our approach and a
demonstration of the advantages of a Bayesian inference-based 
SAM in comparison with the conventional approach. In particular, 
we will show that the common practise of tuning some model parameters 
while keeping others fixed may lead to an incorrect inference 
because of the use of unjustified, strong priors,  
and that our Bayesian inference-based SAM can overcome this problem.  
We will also demonstrate the sensitivity of the inference to 
the error model adopted for the data.
The paper is organised as follows.  In \S\ref{sec:mod}, we describe
our generalised SAM and its relations to other models. A brief
introduction to the principle of Bayesian inference and of the MCMC
technique is presented in \S\ref{sec:met}.  In \S\ref{sec:res_pos},
we show a case study using the stellar mass
function of galaxies as the observational constraint.  The impacts of
prior assumptions and data modelling on the model inference are
presented in \S\ref{sec:res_pri} and \ref{sec:res_cov},
respectively.  Finally, in \S\ref{sec:dis}, we discuss and
summarise our main results.

%%%%%%%%%%%%%%%%%%%%%%%%%%%%%%%%%%%%%%%%%%%%%%%%%%%%%%%%%%%%%%%%%%%%%%%%%%
\section{Semi-analytic model}\label{sec:mod}

As in all other SAMs, our model consists of two main parts, (i) the
assembly of individual dark matter halos, and (ii) gas, radiative and
star-formation processes relevant to galaxy formation.  We first
prepare a large set of halo merger trees with the currently favoured
cosmology and adopt it for all our subsequent semi-analytical modelling of
the baryonic processes.  Since the formation of dark matter halos is
now relatively well understood, we focus on the baryonic physics in
our Bayesian analysis.

\subsection{Halo merger history}

Halo merger trees can either be extracted from cosmological $N$-body
simulations \citep[e.g.][]{Kang2005, Croton2006}, or generated by a
Monte-Carlo method using the extended Press-Schechter formalism
\citep{Lacey1993, Somerville1999, Cole2000, vandenBosch2002}.  Merger
trees from simulations provide the dynamics and environments of the
halo population, but their construction is computationally expensive
and limited by numerical resolution.  On the other hand, Monte-Carlo
merger trees are computationally cheaper to generate and have, in
principal, infinite resolution.  In this paper, we adopt the algorithm
proposed by \citet{Parkinson2008} to generate the merger trees for
halos with a given final ($z=0$) virial mass. This algorithm has been
tuned to match the conditional mass functions found in $N$-body
simulations.  More specifically, as a demonstration we choose the
control parameters $G_0=1$, $\gamma_1=\gamma_2=0$, so that the
resulting halo conditional mass functions are those predicted by the
Extended Press-Schechter conditional mass function
\citep{Parkinson2008}.  We sample a certain number 
of merger trees in each halo mass bin from
$10^{10}\msunh$ to $10^{15}\msunh$, the mass range relevant to the
modelling in this paper.  Since the halos and their merger trees are
randomly sampled from the halo mass function and the conditional mass
function, model predictions based on a finite merger tree sample
suffer from sampling variance.  To reduce such sampling
variance, we generate a sufficiently large number of halo merger trees
in each mass bin so that the variance in model predictions induced
by merger-tree sampling is much smaller than the error in the
observational data used to constrain the model and, hence, can be
ignored. Specifically, we use 1000 merger trees for halos 
with present masses in the range $10^{11}$ - $10^{12.5}\msunh$,  
1500 merger trees in the range $10^{12.5}$ - $10^{13.5}\msunh$, 
400 merger trees in the range $10^{10}$ - $10^{11}\msunh$, and 
about 100 merger trees in the range $10^{13.5}$ - $10^{15}\msunh$. 
Since massive halos are rare in the assumed cosmology, 
their contribution to the scatter of the stellar mass function 
is negligible. We vary the mass resolution of our merger trees
with the final halo mass. For halos with final masses smaller 
than $10^{12}\msunh$, the mass resolution is $10^{9.3}\msunh$; 
for halos with final masses larger than $10^{14}\msunh$, 
it is $10^{11}\msunh$; and for intermediate mass halos, 
it is $10^{10}\msunh$. All the merger trees are sampled
using 60 snapshots equally spaced in $\log(1+z)$ 
from $z=7$ to $z=0$. Throughout the
paper, we use a $\Lambda$CDM cosmology with $\Omega_{\rm M}=0.26$,
$\Obaryon=0.044$, $h=0.71$, $n=0.96$, and $\sigma_8=0.79$, which are
consistent with WMAP5 \citep{Dunkley2009, Komatsu2009}.

\subsection{Radiative cooling}
\label{subs_cooling}

Once the halo formation history is fixed, we model the radiative
cooling of halo gas. As shown in \citet{Lu2010}, the predictions of
often-used cooling models do not agree. Since these models do not
incorporate uncertainties in their cooling prescriptions, the model
choice imposes a strong prior on the SAM.  To compare with
published results, we use the cooling model of \citet{Croton2006}. We will
study the effects of varying the cooling prescription in a future
paper.  In the Croton model, the halo hot gas is redistributed at
every time-step, and the density profile of the hot gas is assumed to
be a singular isothermal profile, \[\rho_{\rm gas}={m_{\rm gas0} \over
  4\pi r_{\rm vir}} r^{-2},\] where $r_{\rm vir}$ is the virial radius
of the halo. The total mass of hot halo gas mass is $m_{\rm gas0}=f_{\rm b}
m_{\rm vir} - \sum_i [m_*^i + m_{\rm cold}^i + m_{\rm out}^i]$, where
$f_{\rm b}=\Omega_{\rm b}/\Omega_0$ is the universal baryon fraction,
$m_*$, $m_{\rm cold}$ and $m_{\rm out}$ are the masses in stars, cold
gas and ejected gas, respectively, and the summation is over all
galaxies in the halo. The temperature of the hot gas is constant for
each halo with $T_{\rm gas} =T_{\rm vir}= 35.9(\frac{v_{\rm vir}}{\rm
  km s^{-1}})^2$K where $v_{\rm vir}$ is the circular velocity of the
halo at the virial radius.  The cooling timescale of the gas at radius
$r$ is then estimated by
\begin{equation}
\tau_{\rm cool}(r)=\frac{3}{2}{\mu m_{\rm H} kT_{\rm gas} \over
  \rho_{\rm gas}(r)\Lambda(T_{\rm gas},Z_{\rm gas})},
\end{equation}
where $\mu$ is the mean molecular weight in units of the mass of
hydrogen atom, and $\Lambda$ is the cooling function from
\citet{Sutherland1993}.  At each time-step, we calculate the cooling
radius $r_{\rm cool}$ by equating the cooling timescale with the
dynamical timescale, $\tau_{\rm cool} =\tau_{\rm dyn} \equiv {r_{\rm
    vir}}/{v_{\rm vir}}$.  If the cooling radius is equal to or
smaller than the virial radius, the cooling rate is defined as
\begin{equation}
\mcooldot
=0.5m_{\rm hot}{r_{\rm cool}\vvir \over r_{\rm vir}^2}.
\end{equation}
In other words, half of
the hot gas mass enclosed by the cooling radius cools and accretes
onto the central object of the halo in a dynamical timescale.  If the
cooling radius is larger than the virial radius, we set the cooling
rate equal to the total hot gas mass in the halo divided by the
dynamical timescale.  We implicitly assume that all hot gas is 
associated with the primary halo and that only the central galaxy can
accrete cooling gas; that is, satellite subhalos contain no hot gas.

In some recent SAMs, Active Galactic Nuclei (AGN) feedback reduces the
gas cooling in massive halos \citep[e.g.][]{Croton2006, Bower2006,
  Somerville2008}.  Equivalently, AGN feedback stops radiative
cooling in halos with masses larger than a characteristic mass 
($\sim 10^{12}\msun$) \citep{Cattaneo2006}.  To include this effect, we
introduce a characteristic halo mass for radiative cooling, 
$M_{\rm CC}$, above which radiative cooling of the hot halo gas is assumed
to be negligible. Since the exact value of $M_{\rm CC}$ is not known a
priori, we treat it as a free parameter in a relatively large mass
range, $10^{11.5}$ - $10^{14.5}\msunh$.

%%%%%%%%%%%%%%%%%%%%%%%%%%%%%%%%%%%%%%%%%%%%%%%%%%%%%%%%%%%%%%%%%%%%%%%%%

\subsection{Star formation}

We assume that the cooled-fraction of halo gas settles into the galaxy
in an exponential disk with scale length $r_{\rm disc}$.  This gas
form stars when the gas disk has a surface density higher than a
certain threshold, $\Sigma_{\rm SF}$, mimicking the critical surface
gas density for star formation seen in disk galaxies
\citep[e.g.][]{Kennicutt1998, Kennicutt2007, Bigiel2008}.  The
fraction of cold gas above the threshold is given by the ratio of the
radius $r_{\rm crit}$ at which the cold gas density is $\Sigma_{\rm
  SF}$ to the disk scale length:
\begin{equation}
r_{\rm crit}/r_{\rm disc} = \ln {m_{\rm cold} \over {2\pi r_{\rm
      disc}^2 \Sigma_{\rm SF}}},
\end{equation}
where $m_{\rm cold}$ is the total cold gas mass of the galaxy.
Therefore, the cold gas mass enclosed by $r_{\rm crit}$ is determined
by the ratio $r_{\rm disc}^2 \Sigma_{\rm SF}/m_{\rm cold}$.
Observationally, the threshold surface density is $\sim 10\msun {\rm
  pc}^{-2}$ \citep[e.g.][]{Martin2001}, although the scale length may
vary.  Theoretically, the disk radius (the scale-length) is related to
the virial radius and the spin parameter of its host halo: $r_{\rm
  disc}\approx { \lambda \over \sqrt{2}}\rvir$ \citep[e.g.][]{Mo1998}.
In cosmological $N$-body simulations, the spin parameters, $\lambda$,
for dark matter halos follow a log-normal distribution with a median of
$\sim 0.05$ \citep[e.g.][]{Warren1992,Cole1996}, but the distribution
of $\lambda$ for the baryonic component that forms galaxy disks is
poorly understood \citep[e.g.][]{Bett2010, Navarro1991}.  In our SAM,
we adopt the fiducial value $\lambda_0=0.05$.  This yields $r_{\rm
  disc,0}=0.035\rvir$ and $\Sigma_{\rm SF,0}=1\msun {\rm pc}^{-2}$.
We then parametrise the term $r_{\rm disc}^2 \Sigma_{\rm SF}=f_{\rm
  SF}r_{\rm disc,0}^2 \Sigma_{\rm SF,0}$.  In the \citet{Croton2006}
model, $r_{\rm disc}$ is set to be $3r_{\rm disc,0}$, and $\Sigma_{\rm
  SF}=10\msun {\rm pc}^{-2}$, so that $f_{\rm SF}=90$.

Using on our parametrisation, the cold gas mass in the disk available
for star formation is
\begin{equation}
m_{\rm sf}=m_{\rm cold}\left[1-\left(1+\ln {m_{\rm cold} \over 2\pi
      f_{\rm SF} \Sigma_{\rm SF,0}r_{\rm disc,0}^2}\right) {2\pi f_{\rm
      SF}\Sigma_{\rm SF,0}r_{\rm disc,0}^2 \over m_{\rm cold}}\right].
\end{equation}
We assume that the star formation rate is proportional to the cold gas
mass within $r_{\rm crit}$ and inversely proportional to the dynamical
timescale of the disk, $\tau_{\rm disc}={r_{\rm disc} \over \vvir}$, 
yielding
\begin{equation}
\dot{m}_*=\epsilon_*{m_{\rm sf} \over \tau_{\rm disc}},
\end{equation}
where $\epsilon_*$ is the star formation efficiency. We assume that
$\epsilon_*$ has a broken power-law dependence on the circular
velocity of the host halo:
\begin{equation}
  \epsilon_* = \left\{ \begin{array}{ll}
      \alpha_{\rm SF} & \mbox{$\vvir \geq V_{\rm SF}$}; \\
      \alpha_{\rm SF}\left({\vvir \over V_{\rm SF}}\right)^{\beta_{\rm SF}}
      & \mbox{$\vvir < V_{\rm SF}$}, \end{array} \right. 
\end{equation}
where $\alpha_{\rm SF}$ and $\beta_{\rm SF}$ are parameters.  Early
models adopted a pure power-law until $\epsilon_*\sim 1$
\citep[e.g.][]{Kang2005}. The \citet{Croton2006} model assumes
$\beta_{\rm SF}=0$ and sets $\alpha_{\rm SF}$ so that 5--15\% of the
cold gas is converted into stars in a disk dynamical time.  The
GALFORM of \citet{Cole2000} considers cases with $\beta_{\rm SF}=0$,
$1.5$ and $2.5$.  In our model, all four parameters, $\alpha_{\rm
  SF}$, $\beta_{\rm SF}$, $V_{\rm SF}$ and $f_{\rm SF}$, are
considered free parameters when modelling star formation in
quiescent disks.
It should be pointed out that our model is 
still based on a specific set of assumptions, even though 
it allows a large range of uncertainties in model parameters. 
There are other prescriptions for star formation that are not 
included in our model \citep[e.g.][]{Somerville2008,Krumholz2009,Fu2010}.

%%%%%%%%%%%%%%%%%%%%%%%%%%%%%%%%%%%%%%%%%%%%%%%%%%%%%%%%%%%%%%%%%%%%%%%%%

\subsection{Supernova feedback}
\label{subs_SNfeedback}

We assume that supernova (SN) feedback affects the interstellar medium
(ISM) and hot halo gas in three ways: (i) the energy feedback from SN
reheats a fraction of the disk ISM from the cold phase to the hot
phase, and the reheated gas is mixed with the hot halo gas; (ii) a
fraction or all of the heated gas is directly ejected from the host
halo without mixing with the hot halo gas; and (iii) if the SN energy
from all galaxies in a halo is sufficiently large, the hot gas in the
host halo can be heated, causing a fraction of the halo hot gas to be
ejected from the halo.  No SAM has incorporated all of these
mechanisms and the strength of the feedback is usually chosen without
strong prior justification.  For example, the Croton model considered
both mechanisms (i) and (iii) \citep{Croton2006}, while GALFORM
incorporated (i) and (ii) \citep{Benson2003c}.  In these models, the
total amount of SN feedback energy is assumed to be related to the
star formation rate, and the feedback is assumed to be instantaneous.
The feedback strength is controlled by a fixed number
\citep[e.g.][]{Croton2006} or assumed to have a power-law dependence
on the circular velocity of the host halo
\citep[e.g.][]{Somerville1999,Cole2000,Kang2005}.  Our model
incorporates all three mechanisms, and their relative strengths are
free parameters.  We assume that for every solar mass of stars formed,
the energy released by supernovae is $\eta_{\rm sn} E_{\rm sn}$, where
$\eta_{\rm sn}$ is determined by the stellar initial mass function
(IMF) and $E_{\rm sn}=10^{51}$erg. Our feedback model enforces energy
conservation, so that the total energy to heat the gas cannot exceed
the total energy released from supernovae.

We write the SN energy released by a mass of $\Delta m_*$ of star
formation as 
\begin{equation}
E_{\rm fb}=\alpha_{\rm SN} \frac{1}{2} \Delta m_* V_{\rm
  SN}^2
\end{equation}
where $V_{\rm SN}=630$km/s and the free parameter $\alpha_{\rm SN}$ 
describes the uncertainties in the feedback energy and in the IMF. For
a Scalo IMF ($\eta_{\rm sn}=5\times 10^{-3}$) and with 20\% of the SN
energy in feedback \citep[e.g.][]{Kang2005}, we find $\alpha_{\rm
  SN}=0.25$. We allow $\alpha_{\rm SN}$ to vary from 0.001 to 10,
encompassing the uncertainty of this parametrisation.  To
conserve energy, the total SN energy released by $m_*$ of star
formation and available for feedback, $E_{\rm fb}$, should be equal to
the sum of the energies used for the reheating, ejection and powering
the wind. Thus, we can write
\begin{equation}
E_{\rm fb}=\frac{1}{2} \left(1-f_{\rm ej}\right) f_{\rm rh} 
    \Delta m_* \vvir^2 +
    \frac{1}{2} f_{\rm ej} f_{\rm rh} \Delta m_* v_{\rm esc}^2 +
    \frac{1}{2} \Delta m_{\rm wind} v_{\rm esc}^2,
\end{equation}
where the coefficients, $f_{\rm rh}$ and $f_{\rm ej}$, control the
mass loading for the reheating and ejection, $v_{\rm esc}$ is the
circular velocity of the current host halo characterising its binding
energy, and $\vvir$ is the circular velocity of the host halo at the
latest time when it was still a primary halo, characterising the
binding energy of the galaxy. Note that $v_{\rm esc}\not=\vvir$ only
for satellite galaxies. We further assume that the fraction for
reheating, $f_{\rm rh}$, has a power-law dependence on the circular
velocity of the halo, $\vvir$. If the galaxy is a satellite, we use
the circular velocity of its host halo when it first became a
subhalo. So we have
\begin{equation}
f_{\rm rh}=\alpha_{\rm RH}\left(\frac{V_0}{\vvir}\right)^{\beta_{\rm RH}},
\end{equation}
where $V_0$ is an arbitrary factor and is set to be $220\,{\rm km/s}$.
The power-law has an upper limit given by energy conservation:
\begin{equation}
f_{\rm rh, max}= \alpha_{\rm SN} \left(\frac{V_{\rm SN}}{\vvir}\right)^2.
\end{equation}
When an amount of $f_{\rm rh} \Delta m_*$ cold gas is reheated, we
assume a fraction $f_{\rm ej}$ escapes from the halo. For simplicity,
we assume $f_{\rm ej}$ has a power-law dependence on the circular
velocity of the current host halo:
\begin{equation}
f_{\rm ej} = \alpha_{\rm EJ}\left(\frac{V_0}{v_{\rm esc}}\right)^{\beta_{\rm EJ}}.
\end{equation}
Again energy conservation sets an upper limit on $f_{\rm ej}$:
\begin{equation}
  f_{\rm ej, max}= \left[\frac{f_{\rm rh, max}}{f_{\rm rh}} - 1\right]
  \times \left[\left(\frac{v_{\rm esc}}{\vvir}\right)^2-1\right]^{-1}.
\end{equation}
If there is still energy available after reheating and ejection, the
surplus is assumed to power a wind, and the mass of the wind can be
written as
\begin{equation}
\Delta m_{\rm wind} = \epsilon_{\rm W} \Delta m_* 
  \left\{ \alpha_{\rm SN}\left(\frac{V_{\rm SN}}{v_{\rm esc}}\right)^2 -
  f_{\rm rh} \left[\left(\frac{\vvir}{v_{\rm esc}}\right)^2+f_{\rm ej}\right]\right\}.
\end{equation}
We assume that a fraction of $f_{\rm RI}$ of the gas in the outflow,
ejection and wind will come back to the halo as hot gas in a dynamical
timescale, and we treat $f_{\rm RI}$ as a free parameter.

Thus, we model the SN feedback with 7 parameters: $\alpha_{\rm SN}$,
$\alpha_{\rm RH}$, $\beta_{\rm RH}$, $\alpha_{\rm EJ}$, $\beta_{\rm
  EJ}$, $\epsilon_{\rm W}$ and $f_{\rm RI}$.  Because the wind
dominates the outflow, we find that $\alpha_{\rm EJ}$ and $\beta_{\rm
  EJ}$ are not constrained by the stellar mass function alone. 
Therefore, we fix $\alpha_{\rm EJ}=0$ and
$\beta_{\rm EJ}=0$ in the present paper.
Our model shares a number of common parameterisations 
with other models. For example, the reheating model is similar to 
the model studied in \citet{Bower2010}; if $\alpha_{\rm EJ}$ and $\beta_{\rm RH}$
are set to be 0, our model is reduced to the Croton model \citep{Croton2006}; 
if $\epsilon_{\rm W}$ is set to be 0, our model is similar to the model 
described in \citet{Somerville2008}. However, it is worth pointing out 
that other parameterisations are also possible \citep[e.g.][]{Benson2003c}. 
 
%%%%%%%%%%%%%%%%%%%%%%%%%%%%%%%%%%%%%%%%%%%%%%%%%%%%%%%%%%%%%%%%%%%%%%%%
\subsection{Galaxy mergers}

When two dark matter halos merge, we simply add the dark matter and
hot gas of the smaller halo to the bigger one. The central galaxy of
the more massive halo is then treated as the central galaxy of the new
halo, and all other galaxies are considered as satellites.  A
satellite galaxy merges with the central galaxy in some fraction
$f_{\rm DF}$ of the dynamical friction timescale.  The dynamical
friction timescale is parametrised as
\begin{equation}
t_{\rm fric}={1.17 \rvir^2 \vvir \over \ln \Lambda G M_{\rm sat}}, 
\end{equation}
where $\rvir$ and $\vvir$ are the virial radius and circular velocity
of the new host halo, $M_{\rm sat}$ is the mass of the previous host
halo of the satellite before it merges into the current halo, and $\ln
\Lambda$ is the Coulomb logarithm, which is modelled as $\ln \Lambda =
\ln (1+M_{\rm vir}/M_{\rm sat})$ \citep[e.g.][]{Croton2006}.  This
formula assumes that the satellite galaxy is hosted by a subhalo with
mass $M_{\rm sat}$ and orbits in a central halo with a singular
isothermal density profile of circular velocity $\vvir$, starting at the
virial radius \citep{Binney1987}.  Earlier SAMs adopted similar
parameterisations, but used different prefactors. For example, some
SAMs chose the galaxy mass for $M_{\rm sat}$ \citep[e.g.][]{Cole2000}
and some others chose the subhalo mass for $M_{\rm sat}$
\citep[e.g.][]{Croton2006}; this results in an order of magnitude
difference in the prefactor.  Other uncertainties include the value of
the Coulomb logarithm, the effect of tidal stripping on orbital decay,
and the initial velocity of the satellite. In our model, these
uncertainties are absorbed into the prefactor, $f_{\rm DF}$, a free
parameter.

The merging timescale is calculated when the host halo of the
satellite merges into the host halo of the central galaxy. If the
satellite was already a satellite before the merger, the dynamical
fraction timescale for the satellite is recalculated based on the
properties of the new host.  When a satellite galaxy merges into the
central galaxy, our treatment for the merger remnant depends on the
mass ratio of the two galaxies, $m_{\rm sat}/m_{\rm central}$. Mergers
are considered as major or minor depending on whether
$m_{\rm sat}/m_{\rm central}$ is larger or smaller than a
pre-selected $f_{\rm MG}<1$. The values of $f_{\rm MG}$ adopted in
earlier SAMs are $\sim 0.3$. As the choice of this parameter is not
constrained by the stellar mass function of galaxies, we simply take
$f_{\rm MG}=0.3$ instead of treating it as a free parameter.

For a minor merger ($m_{\rm sat}/m_{\rm central} \leq 0.3 $), the
satellite's stars are added to the central bulge, and the satellite's
gas is added to the central disk. A minor merger is assumed to trigger
a star-burst in the disk, and all the stars formed in the burst are
added to the disk component.  For a major merger ($m_{\rm sat}/m_{\rm
  central} > 0.3$), we combine all the existing stars from the two
merging galaxies into a central galaxy, which is now assumed to be an
elliptical.  Each major merger triggers a star-burst, and all stars
formed in the burst are added into the central elliptical galaxy. A
fraction $e_{\rm burst}$ of the combined cold gas in the two merging
progenitors becomes stars, and the rest joins the gaseous disk. We
assume that $e_{\rm burst}$ depends on the ratio of the baryon masses
of the two galaxies: $e_{\rm burst}=\alpha_{\rm burst}(m_{\rm
  sat}/m_{\rm central})^{\beta_{\rm burst}}$.

Similar models for galaxy mergers were adopted by
\citet{Somerville2001, Somerville2008} and \citet{Croton2006} although
different authors used different values for the model parameters.  In our
model, the four parameters in the parametrisation, $f_{\rm DF}$,
$f_{\rm MG}$, $\alpha_{\rm burst}$ and $\beta_{\rm burst}$, are all
treated as free parameters.  As mentioned above, 
since $f_{\rm MG}$ is not constrained by
the stellar mass function considered in this paper, we simply fix its
value to be $0.3$.

\subsection{Calculation of a single model}

\begin{figure}
  \centering
  \includegraphics[width=0.8\textwidth, height=0.8\textheight]{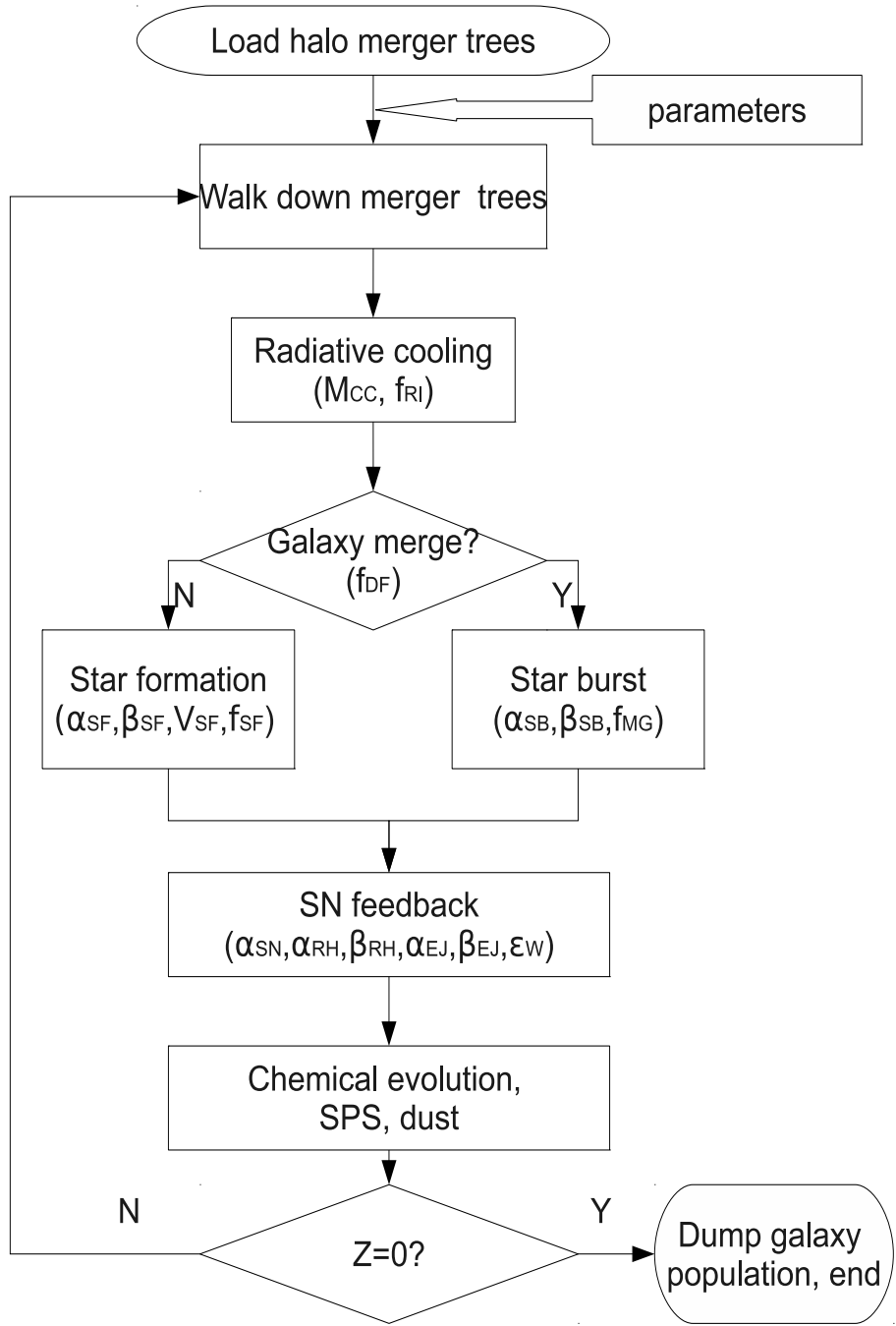}
  \caption {Flow chart describing the calculation of our semi-analytic
    model of galaxy formation. The parameters explored in the present
    paper are listed in the corresponding blocks.}
  \label{fig:sam}
\end{figure}

The flowchart shown in Figure \ref{fig:sam} summarises the calculation
of the SAM described above.  The code loads merger trees and other
tables (e.g. cooling functions, stellar mass-to-light ratios for
different star formation histories, and dust extinction) for
subsequent calculations, and then reads the model parameters
introduced above in this section, which are summarised in Table 1.  We walk each
merger tree from the top (the initial time) to bottom
(the present time). At each tree level, a galaxy grows in the centre of a
halo if the halo does not have any progenitor halos.  If the halo is
assembled through the mergers of progenitor halos, the central galaxy
of the most massive progenitor is considered to be the central galaxy
of the current halo, and all the other existing galaxies are considered to
be satellites.

At the initial time, we distribute hot gas in the dark matter halos
and radiative cooling begins. We sub-divide each of the 60 
time steps used to sample a merger tree into 5 finer time 
steps (equally spaced in $t$)
to compute the cooling and to evolve the galaxies.
In every time step, gas that is able to cool in the current time step is
assigned to the central galaxy.  For all galaxies in the halo, star
formation continues until the cold gas 
surface density drops below the threshold value. When a satellite
galaxy merges into a central galaxy, the recipes for the morphological
transformation and merger-triggered starburst are applied. For both
star formation modes, quiescent or bursts, we calculate the
effects of SN feedback. We model chemical evolution in the ISM using the
``instantaneous recycling approximation'' \citep[][]{Cole2000}: a
fraction $R$ of the newly formed stellar mass and a yield $p$ of heavy
elements are instantaneously returned to and uniformly mixed with the
cold gas. Metals enrich the halo gas as the reheated gas mixes with
the hot halo gas (assuming a one-zone model, see Subsection
\ref{subs_SNfeedback}) and affect the cooling rate.  Both $R$ and $p$
depend on the IMF. However, since we have adopted a simplified model for gas
cooling (see Subsection \ref{subs_cooling}) and since the stellar mass
function we are concerned with here is affected by metallicity only through
gas cooling, in this paper we simply fix $R=0.3$ and $p=0.03$ instead
of treating them as free parameters.  Our code uses the Stellar
Population Synthesis (SPS) model of \citet{Bruzual2003} and the dust
model of \citet{Kauffmann1999} to assign fluxes to galaxies.

The evolution continues until the root of the merger tree is reached.
At this point, we have a realisation of the model specified by the set
of parameters. The results obtained from these realisations can then
be used to compare with observational data to evaluate the likelihood
of the data given the model.

%%%%%%%%%%%%%%%%%%%%%%%%%%%%%%%%%%%%%%%%%%%%%%%%%%%%%%%%%%%%%%%%%%%%%%%%%%
\begin{table}
%\caption{Model parameters} % title name of the table
\centering
\begin{tabular}{l c c c c c}
\hline\hline
\# & Parameter & Meaning & Prior & Posterior
\\ [0.5ex]
\hline
 &                                  &  & [1.5, 4.5] & [2.19, 2.67] [3.09, 4.47] \\
1 & $\log M_{\rm CC} (\msun) $ & cooling cut-off halo mass& [1.5 , 4.5] & [2.13, 2.49] [3.15, 4.47] \\
 &                                  &  & [1.5, 4.5] & [2.07, 2.49] \\
\hline
 &                                           &  & [-3, 0] & [-2.19, -0.03] \\
2 & $\log \alpha_{\rm SF}$ & star formation efficiency power-law amplitude & [-3, 0] & [-2.97, -2.85] [-1.47, -0.03] \\
 &                                           &  & [-3, 0] & [-2.97, -2.49] \\
\hline
 &                                                       &  & [-1, 12] & [-0.87, 0.43] [3.29, 11.87] \\
3 & $\beta_{\rm SF}$ & star formation efficiency power-law index &  [-0.2, 0.2] & [-0.2, 0.2] \\
 & &  & [-0.2, 0.2] & [-0.2, 0.2] \\
\hline

 &  & & [1.5, 3.0] & [1.52, 2.39] \\
4 & $\log V_{\rm SF} $ (km/s) & star formation law turn-over halo circular velocity & [2.1, 2.3] & [2.1, 2.3] \\
 &  & & [2.1, 2.3] & [2.1, 2.3] \\
\hline

 &  & & [-1, 3] & [-0.96, -0.64] [-0.24, 2.16] \\
5 & $\log f_{\rm SF} (\msun/{\rm pc}^2)$ & star formation threshold gas surface density & [1.8, 2.2] & [1.8, 2.2]  \\
 &  & & [1.8, 2.2] & [1.8, 2.2] \\
\hline

 & &  & [-3, 1] &  [-2.35, 0.85] \\
6 & $\log \alpha_{\rm SN}$ & SN feedback energy fraction & [-3, 1] & [-1.35, -1.15] [-0.25, 1.00] \\
 & &  & [-3, 1] & [-1.75, 0.25] \\
\hline

 & &  & [-3, 2] & [-2.55, 1.95] \\
7 & $\log \alpha_{\rm RH}$ & SN feedback reheating power-law amplitude & [-3, 2] & [-2.65, -0.75] \\
 & &  & [-3, 2] & [0.260, 1.22] [-0.75, 1.95]\\
\hline

 & &  & [0, 14] & [0.14, 13.86] \\
8 & $\beta_{\rm RH}$ & SN feedback reheating power-law index & [0, 14] & [5.46, 11.62] \\
 & &  & [1.8, 2.2] & [1.8, 2.2] \\
\hline

 & &  & [-3, 0] & [-2.97, -0.15] \\
9 & $\log \epsilon_{\rm W}$ & fraction of surplus SN feedback energy used for powering wind & [-3, 0] & [-2.97, -0.81] \\
 & &  & [-3, 0] & [-2.97, -0.21] \\
\hline

 & &  & [-2, 0] & [-1.98, -0.02] \\
10 & $\log f_{\rm RI}$ & fraction of re-infall ejected hot gas & [-2, 0] & [-1.97, -0.03] \\
 & &  & [-2, 0] & [-1.94, -0.02] \\
\hline

 & &  & [-1, 2] & [0.53, 1.97] \\
11 & $\log f_{\rm DF}$ & merging time-scale in dynamical friction time-scale & [-1, 2] & [0.23, 0.59] [0.77, 1.97] \\
 & &  & [-1, 2] & [0.05, 0.65] \\
\hline

 & &  & [-2, 0] & [-1.98, -0.02] \\
12 & $\log \alpha_{\rm SB}$ & merger triggered star burst efficiency power-law amplitude& [-2, 0] & [-1.97, -0.09] \\
 & &  & [-2, 0] & [-1.97, -0.15] \\
\hline

 & &  & [0, 2] & [0.02, 1.98] \\
13 & $\beta_{\rm SB}$ & merger triggered star burst efficiency power-law index & [0, 2] & [0.02, 1.98] \\
 & &  & [0, 2] & [0.02, 1.98] \\
\hline
\hline

 & & & & \\
14 & $\alpha_{\rm EJ}$ (fixed) & SN feedback cold gas ejection power-law amplitude & 0.0 & 0.0 \\
 & & & & \\
\hline
 & & & & \\
15 & $\beta_{\rm EJ}$ (fixed) & SN feedback cold gas ejection power-law index & 0.0 &  0.0 \\
 & & & & \\
\hline
 & & & & \\
16 & $f_{\rm MG}$ (fixed) & major merger minor merger threshold & 0.3 & 0.3 \\
 & & & & \\
\hline

\end{tabular}
\caption{
Summary of the model parameters. 
Column 2: the parameter; Column 3: the description of the parameter; 
Column 4: the prior distribution; Column 5: the 95\% confidence bound of 
the posterior distribution. 
For the prior and posterior distributions, 
the three rows for each parameter are for Case 0, Case 1 and Case 2, respectively. 
Parameter 1 to 13 are set free, whereas parameter 14, 15 and 16 are fixed.
}
\label{tab:par}
\end{table}

%%%%%%%%%%%%%%%%%%%%%%%%%%%%%%%%%%%%%%%%%%%%%%%%%%%%%%%%%%%%%%%%%%%%%%%%%%
\section{Bayesian model inference and the MCMC method}\label{sec:met}

\subsection{Bayesian inference}

Bayes Theorem states that the posterior probability of a set of
parameters $\mathbf{\Theta}$ in a model (or hypothesis) $H$ for given
data $\mathbf{D}$ is
\begin{equation}
  P(\mathbf{\Theta}|\mathbf{D},H) \propto  P(\mathbf{\Theta}|H)
  L(\mathbf{D}|\mathbf{\Theta}, H),  
\end{equation}
where $P(\mathbf{\Theta}|H)$ is the prior probability distribution, which
describes any knowledge acquired about the parameters before seeing
the data, and $L({\bf D}|\mathbf{\Theta}, H)$ is the likelihood of the
data ${\bf D}$ for the given model parameter set $\mathbf{\Theta}$. 
As mentioned earlier, for SAMs, we have limited prior
knowledge about the model parameters.  Therefore, we choose
either uniform or $1/x$ distributions between two physically
chosen bounds for the prior distributions, depending on the particular
parameter in question. As a test, the
priors for some of the parameters are made strongly restrictive to
demonstrate the sensitivity to these choices.  Our assumed priors
for the standard model (Case 0) are summarised in Table \ref{tab:par} as the
first listed for each parameter. 
Note that the prior adopted for $\alpha_{\rm SN}$ allows the model to 
use more energy to power the feedback than the total SN energy
assumed to be available. With such a prior, we test 
whether the model could explain the data if the SN energy 
is somehow underestimated.
The problem-specific definition
of the likelihood function is described in later sections.

\subsection{The Markov-Chain Monte-Carlo algorithm}

Since it is not possible to integrate the posterior probability distribution
function analytically for our SAM,  we use a Monte-Carlo
sampling approach to elucidate the posterior distribution.  We adopt a
newly developed software package, the Bayesian Inference Engine
\citep[BIE,][]{BIE2010,Weinberg2010}, which includes a suite of 
advanced MCMC algorithms and supports parallel computation. 
A detailed description of the package is beyond the scope of 
the present paper and can be found in the two references cited above. 
Here we present a brief description.  

As we will show later, the topological structure of the posterior
probability distribution in our problem is high-dimensional and very
complex. Not only does the posterior show multi-modality and strong
degeneracies among the model parameters, but also the high-probability
regions only occupy a very small fraction of the entire parameter
space along a curving, very thin manifold \citep[also see][]{Bower2010}.
Because of this, it is technically challenging to sample the
posterior efficiently using the standard Metropolis-Hastings MCMC
algorithm. To overcome this problem, we adopt differential 
evolution algorithm as the main algorithm for our MCMC sampler 
\citep{TerBraak2006}. 
For every single chain at each step, the differential evolution 
algorithm randomly selects two other chains and uses a fraction 
of the vector connecting the current states of the two chains as 
a proposal. This strategy improves proposal efficiency and mixing 
by automatically ``tuning'' the proposals to the ensemble of states 
comprising the individual chains.
For a multi-dimensional Gaussian posterior, the optimised fraction of
the vector is $\gamma_0=2.38/\sqrt N_{\rm dim}$, where $N_{\rm dim}$
is the dimension of the parameter space \citep{TerBraak2006}.  Since
our posterior is expected to deviate significantly from a Gaussian, we
use $\gamma=0.1\gamma_0$ to maintain a good acceptance rate ($\approx
10\%$). After every 10 steps, we use the full vector as the proposal by
temporally setting $\gamma=1$ to allow the chains to swap modes.  As the
simulation proceeds, the chains gradually settle into the high
probability regions and the distribution can guide the chains to move
along the ridges of the posterior or to jump between different
modes. Moreover, all the converged chains sample the posterior, further
enhancing the overall efficiency.  

To enhance mixing and to explore the parameter space more efficiently,  
we combine a tempered simulation algorithm \citep{Neal1996} 
with differential evolution. In short, tempered
simulation proposes exchanges between the posterior distribution of
$P_0$ and a ``powered-up'' distribution $P_j\propto P_0^{1/T_i}$ with
$T_i\le T_{\rm max}$.  Each step begins by ``melting'', $T_{i+1}>T_i$
followed by ``freezing'', $T_{i+1}<T_i$.  We perform one tempered
step for every 21 standard steps, with the maximum temperature 
$T_{\rm max}$ selected to be similar to the difference in the 
logarithmic posterior probability between a high-probability region and a
low-probability valley: $T_{\rm max} \approx \ln P_{\rm max}/P_{\rm min}$. In
the temperature range from 1 to $T_{\rm max}$, we set $M$ temperature
levels equally spaced logarithmically. The default value of $M$
is set to be $\sqrt{N_{\rm dim}+3} \ln T_{\rm max}$. For our problem,
$N_{\rm dim}=13$ and we set $T_{\rm max}=64$, so that $M=16$. At each
temperature level $T_i$, 10 differential evolution steps are taken,
with $\gamma$ stretched by a factor of $T_i^{1/2}$. In total, it takes
320 differential evolution steps for a chain to go through the
``melting'' and ``freezing'' procedure for a single tempered
step. As the parameter $T_{\rm max}$ controls the efficiency 
with which the MCMC chains explore parameter space,
we have carried out tests by varying 
the maximum temperature. The tests show that the $T_{\rm max}$ 
we choose is sufficiently high: the posterior does not show 
any new features as the temperature is increased by a factor of 4. 
Using a suite of tests, we also find that the tempered steps 
with the $T_{\rm max}$ specified as above  substantially 
improves the efficiency of exploring state space, 
speeding up convergence, although employing such steps significantly
increases the computational load.

\subsection{A Bayesian-inference based SAM }

\begin{figure}
  \centering
  \includegraphics[width=0.8\textwidth, angle=0.0]{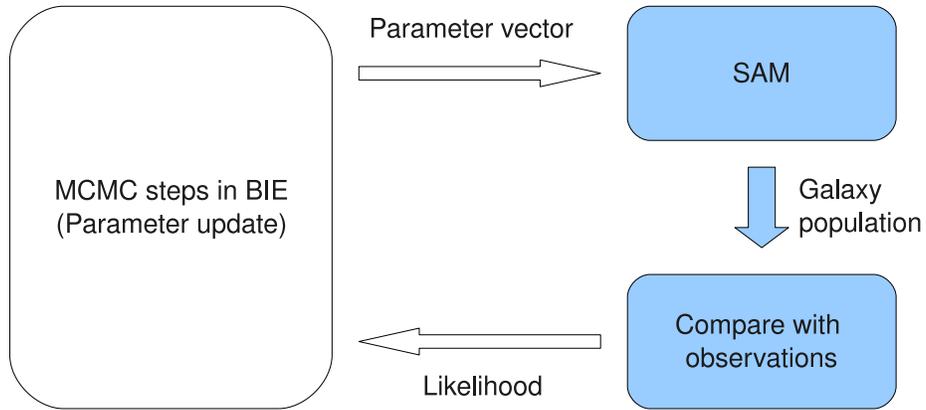}
  \caption {A flow chart describing the structure of our
    Bayesian-inference based semi-analytic model.}
  \label{fig:bie_sam}
\end{figure}

We outline the structure of our Bayesian-inference based SAM in
Figure \ref{fig:bie_sam}. The MCMC algorithm provides proposal
parameter vectors for the SAM, and the SAM predicts the galaxy
population using the proposed parameter set.  The likelihood is
evaluated by comparing the model prediction with the data, and is returned
to the MCMC.  The MCMC algorithm accepts or rejects the proposal
based on the posterior probability and generates a new proposal for
the SAM.
We use our tempered differential evolution algorithm described above with
128 chains running in parallel.
The MCMC-SAM loop continues until convergence is
achieved. The convergence of the chains is monitored by the
Gelman-Rubin ${\hat R}$ statistic \citep{Gelman1992}.  In essence,
${\hat R}$ is the ratio of the variance between chains to the variance
within chains.  We declare convergence when ${\hat R}\leq 1.2$.  When
the chains are converged, we use post-convergence states (typically
about $10^6$) to study and characterise the posterior distribution.
The converged states sample the full probability distribution of the
model parameters given the observational data, and can be used to estimate
confidence regions for individual model-data comparisons through
marginalisation and to determine the relative posterior odds for
different models.  In the following sections, we use a simple example
to demonstrate the power of our Bayesian-inference based SAM.

%%%%%%%%%%%%%%%%%%%%%%%%%%%%%%%%%%%%%%%%%%%%%%%%%%%%%%%%%%%%%%%%%%%%%%%%%%
\section{The SAM Posterior: Stellar Mass Function Constrained}
\label{sec:res_pos}

In this paper, we consider constraints on our SAM provided by the
stellar mass function of galaxies, a fundamental property of the
galaxy population that has been extensively used for model--data
comparison. We choose the stellar mass function instead of the
luminosity function simply because the stellar mass of galaxies is a
direct prediction of our SAM, and hence we avoid problems associated with any
uncertainties in the stellar population synthesis or dust models in our
predictions.  However, these same uncertainties are
present in the reduced data, since a stellar population synthesis
model was used to convert the observed galaxy luminosities into
stellar masses. These uncertainties should in principle be properly
included in the error budget of the observational data. 
Since the purpose of this section is purely to provide a concrete 
demonstration of our method and to illustrate the complexities inherent in
the posterior distribution function, we adopt an ad hoc model for the errors.
In \S\ref{sec:res_cov}, we explore the impact of the error 
model on the Bayesian inference by changing our assumptions about the errors. 

We study the constraints on the 13 free parameters characterising our
SAM (see Table 1) using the stellar mass function of \citet{Bell2003}.
Assuming that stellar mass bins are mutually independent, the
likelihood function is
\begin{equation}\label{eq:lik_ind}
  L(\Phi_{\rm obs}|\theta)=L_0 \exp 
  \left\{-\sum_i{\left[\Phi_{i,{\rm obs}}
        -\Phi_{i,{\rm mod}}(\mathbf{\theta})\right]^2 
      \over 2 \sigma_{i,{\rm obs}}^2}\right\},
\end{equation}
where $L_{0}$ is an arbitrary normalisation factor, $\Phi_{i,{\rm
    obs}}$ is the value of the observed stellar mass function in the
$i$th bin, $\Phi_{i,{\rm mod}}$ is the corresponding value predicted
by the model with the given parameter set $\mathbf{\theta}$, and
$\sigma_{i, {\rm obs}}^2$ is the variance of the observed stellar mass
function. The error estimation of \citet{Bell2003} only takes into
account the sampling error, but we expect significant bias (systematic
uncertainty) from the assumptions made in the data reduction.  To mimic
the effect of large systematic uncertainty, we artificially inflate
the statistical error bars by a factor of 3.  Please note, we are not
advocating this procedure, rather, we argue this is a \emph{very bad}
thing to do in general for at least two reasons: (i) this tends to
imply greater support for a model than is truly admitted by the data,
and conversely, tends to reduce the ability of the data to choose
between competing hypotheses; and (ii) inflated error may hide serious
problems with the data or inconsistencies with other data.  Strictly
speaking, the Bayesian approach applies equally well to systematic
uncertainty as to sampling error.  Mathematically, let systematic
uncertainties be described by a parameter vector $\mathbf{\eta}$.  The
likelihood now depends on $\mathbf{\eta}$ through $\Phi_{i,{\rm
    obs}}(\mathbf{\eta})$.  We simply define a prior distribution for
the uncertainty $P(\mathbf{\eta})$ by expert opinion or through an
ancillary calibration.  The inference continues as before, now with
the augmented parameter vector $\mathbf{\Theta} = (\mathbf{\theta},
\mathbf{\eta})$.  In the end, we simply marginalise over
$\mathbf{\eta}$.  For our problem specifically, we are aware our error
inflation produce is ad hoc and does not correctly represent the
bin-to-bin covariance in $\Phi_{i,{\rm obs}}$ induced by the stellar
mass function.  We will discuss how such covariance affects our
results in \S\ref{sec:res_cov}.  We will perform a luminosity
function-based inference using a population synthesis model and an
appropriately chosen prior uncertainty in a future paper.  However,
the lack of a stellar mass function with a suitably described error
model forces us to make a crude error model approximation for the
point of illustration in this section.  In addition, our
Monte-Carlo evaluation of $\Phi_{i,{\rm mod}}$ has variance 
owing to the finite sampling of the assembly histories of dark matter halos. 
This model dispersion should be included in the likelihood. 
However, it is typically 3 times smaller than the inflated 
error bars in the data and, therefore, not explicitly included 
in equation (\ref{eq:lik_ind}).

Our model also has other uncertainties. For example, the 
cooling rate given by different prescriptions can vary by 
a factor of a few \citep[see][]{Lu2010}. Such uncertainties 
could be included in the likelihood evaluation if they were 
properly understood \citep[e.g.][]{Bower2010}.   
Alternatively, one may include model uncertainties as 
part of the model by using an extended model family.
In this paper, we restrict our demonstration to a fixed 
model family and ignore any uncertainties other than those
represented by the priors. In a future paper we will 
demonstrate how to extend the analysis to multiple model 
families using Bayesian model selection.

\begin{figure}
  \centering
  \includegraphics[width=0.8\textwidth, angle=0.0]{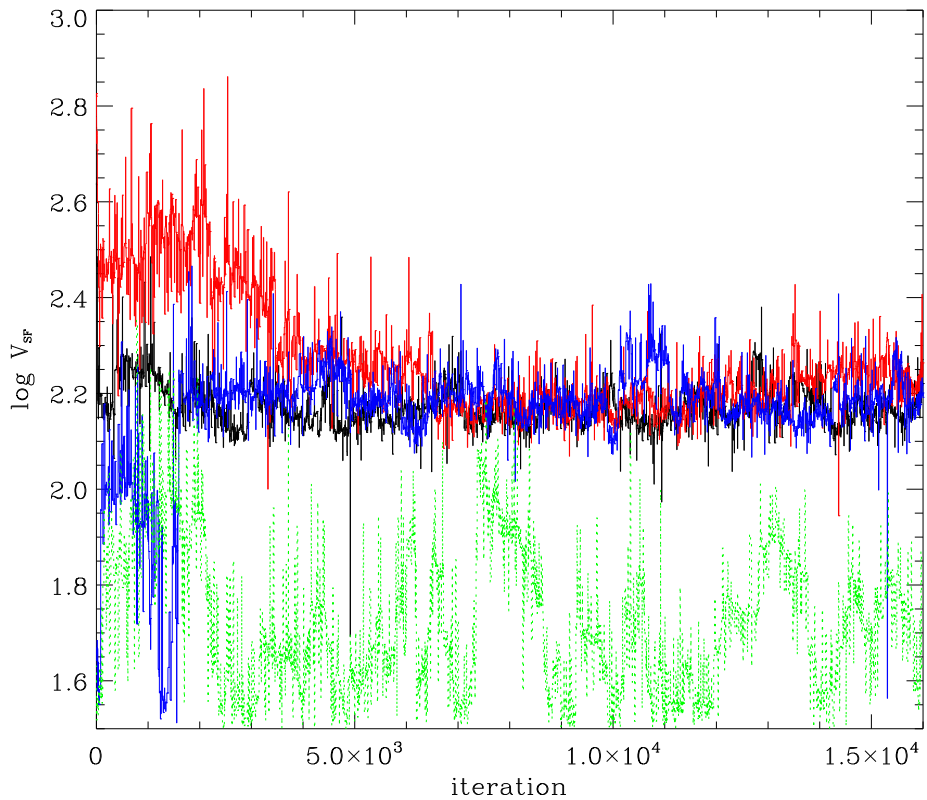}
  \caption {The trajectories of four MCMC chains run with the Tempered 
	Differential Evolution algorithm in the dimension of the parameter 
	$V_{\rm SF}$. The three solid lines (black, red and blue) are three 
	randomly selected converged chains, while the dotted line (green)
	is an outlier chain as identified by the Gelman-Rubin statistic. 
	}
  \label{fig:chains}
\end{figure}

We use our Tempered-Differential Evolution algorithm to run 
the MCMC simulation with 128 chains in parallel. The initial states 
of the chains are randomly distributed in parameter space according 
to the prior probability distribution. We terminate the simulation after
16,000 iterations, when a sufficiently large number of 
states are collected to summarise the marginalised posterior. 
The Gelman-Rubin statistic monitors the convergence of the MCMC simulation, 
and it identifies 123 chains that are well mixed after the simulation terminates.
In Figure \ref{fig:chains}, we plot 
the trajectories of 3 chains randomly selected from the mixed 
chains and compare them with a trajectory of a outlier chain. 
One sees that the chains were all widely dispersed at the 
beginning. The mixed chains gradually converge to a 
high probability mode after about 3000 iterations.
In contrast, the outlier chain does not converge, but 
wanders around in low probability regions. The 
simulations are kept running for 16,000 iterations, 
even though most of the chains have ``burned-in'' 
after 4000 iterations. For the analysis presented in this 
section, we take the consecutive 12,000 steps of the 123 
converged chains, about 1.5 million states, to summarise the 
marginalised posterior probability distributions of 
the model parameters.  

\subsection{Physical implications}

Figure \ref{fig:par_mod0} shows the one- and two-dimensional
marginalised posterior probability distributions of the 13 free
parameters.  Three of these parameters, $f_{\rm RI}$, $\alpha_{\rm
  SB}$ and $\beta_{\rm SB}$, are unconstrained by the stellar mass
function and not shown.  In the upper-right corner of the figure, we
plot the predicted stellar mass function by marginalising over the
95\% confidence range of the posterior.  Clearly, the stellar mass
function is well reproduced by the model.  Table \ref{tab:par} lists
the 95\% confidence bounds of all the parameters (as the first
range listed for each parameter).

The strength of the constraints varies widely.  Some parameters are
weakly constrained: for example, $\epsilon_{\rm W}$, the efficiency of SN
feedback powering the galactic wind, is very weakly constrained.  In
contrast, some parameters are tightly constrained: for example,
$V_{\rm SF}$ is constrained to a narrow range (around $\sim160$km/s),
so are $\beta_{\rm SF}$ (around $6$) and $\beta_{\rm RH}$ (around
$8$).  Our inferred values of $\beta_{\rm SF}$ and $\beta_{\rm RH}$
are much higher than those adopted in previous SAMs.  The
posterior indicates a sharply suppressed star formation efficiency in
halos with circular velocities below $\sim160$km/s.  In addition, the
posterior distribution in the $\beta_{\rm SF}$ - $\beta_{\rm RH}$
plane reveals bimodality: either the star formation efficiency or the
SN reheating efficiency is a steep power-law of halo circular
velocity.  In other words, the shallow slope of the stellar mass
function at the low-mass end requires the suppression of star
formation in small halos.  Since the star formation efficiency directly
controls the conversion of cold gas into stellar mass, the
high $\beta_{\rm SF}$ mode dominates the high $\beta_{\rm RH}$ mode.
We are unsure whether or not such high values of $\beta_{\rm SF}$ and 
$\beta_{\rm RH}$ are physically plausible. It is likely that some new physics
in addition to SN feedback is required to suppress star formation in
low-mass halos, as we will demonstrate in detail in a forthcoming
paper.  

Some model parameters are strongly correlated. These include the
following pairs of parameters: $f_{\rm SF}$--$\alpha_{\rm SF}$;
$\alpha_{\rm RH}$--$\beta_{\rm RH}$; and $M_{\rm CC}$--$f_{\rm DF}$.
Both $\alpha_{\rm SF}$ and $f_{\rm SF}$ control the conversion of cold
gas into stars and the degeneracy is expected.  Similarly, the two parameters in
the power-law parametrisation of the SN reheating, $\alpha_{\rm RH}$
and $\beta_{\rm RH}$, are degenerate.  And again, the parameters
controlling the two mechanisms responsible for the formation of
central galaxies in massive halos, $M_{\rm CC}$ and $f_{\rm DF}$ are
correlated; massive central galaxies can either acquire their mass
through gas cooling and {\it in situ} star formation, or through the
accretion of satellite galaxies. The observed sharp decline of the
stellar mass function at the high-mass end requires either that
gas cooling in halos more massive than $\sim10^{12}\Msun$ is effectively
quenched (e.g. by AGN feedback) or that the merger rate of satellite galaxies
into the central galaxy by dynamical friction is slow.

Comparing our results with those of previous studies, 
we notice that some of the modes we identified are broadly 
consistent with those found by other studies. \citet{Henriques2009} 
found that $\epsilon_{\rm disc}$ in the model proposed by \citet{Croton2006},  
which corresponds to $\alpha_{\rm RH}$ in our model, 
is required to be as high as about 10. \citet{Bower2010} found 
that the normalisation for the star formation efficiency 
is as low as about 0.003, which is also similar to the mode 
we find for $\alpha_{\rm SF}$. Nevertheless, the features shown 
in the posterior distributions in these studies and ours do not 
generally agree because of different definitions of the parameters. 

\subsection{Structure of the posterior distribution}

\begin{figure}
  \centering
  \includegraphics[width=\textwidth]{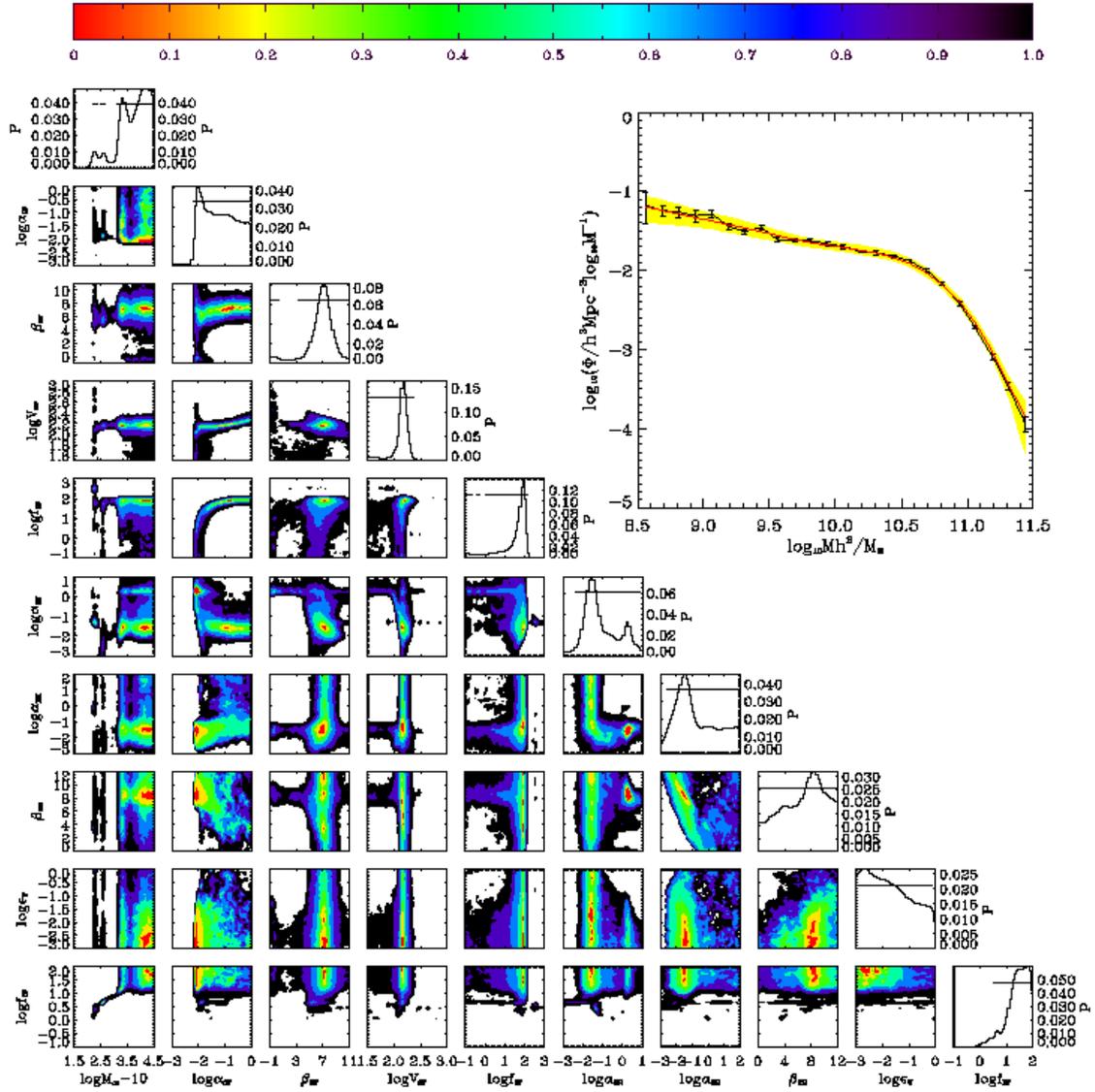}
  \caption {The marginalised posterior distribution for key parameters
    for our fiducial run (Case 0). The colour coding represents
    confidence levels as shown by the colour-bar on the top of the
    figure.  The horizontal bars in the one-dimensional marginals
    indicate the 95\% confidence interval.  The observed stellar mass
    function of galaxies (black line and error bars) from
    \citep{Bell2003} together with the marginalised model prediction
    is inset. The red solid line shows the median value of the
    prediction, while the yellow shaded region represents the 95\%
    confidence interval. }
  \label{fig:par_mod0}
\end{figure}

\begin{figure}
  \centering
\includegraphics[width=\textwidth]{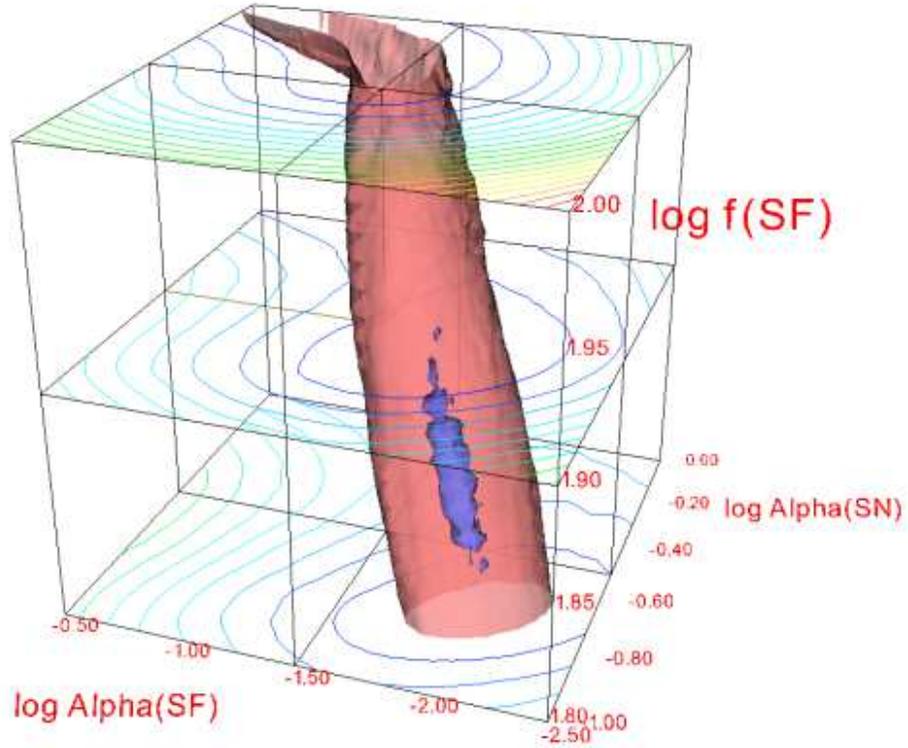}
  \caption{The likelihood function in a three-dimensional space defined by 
	$\log \alpha_{\rm SF}$, $\log \alpha_{\rm SN}$ and 
    $\log f_{\rm SF}$.
	Other parameters are fixed at the values where the posterior peaks.
	The inner surface (blue) has $\log(L)=-4$, and the outer 
	surface (red) has $\log(L) =-9.9$ (so the latter value is approximately 
	``1-$sigma$'' if the likelihood function were Gaussian). 
	The contour lines on the planes are linearly spaced in the logarithmic scale 
	with a spacing of 13.4. 
    This figure demonstrates the complexity of the likelihood function 
	and the posterior distribution. 
  }\label{fig:pos_cut}
\end{figure}

The two-dimensional posterior distributions shown in Figure
\ref{fig:par_mod0} are marginalised over 11 dimensions and wash out
much of the intrinsic sub-dimensional structure that complicates the
inference and renders tweaking by hand unreliable. To demonstrate
this, Figure \ref{fig:pos_cut} shows a 3-dimensional cut through
the 13-dimensional likelihood function. The cut is made by computing 
the likelihood function on a fine grid of $\alpha_{\rm SF}$, 
$\alpha_{\rm RH}$ and $f_{\rm SF}$ and fixing the other 
10 parameters to the values where the likelihood function has its
global maximum. In the plotted volume, the maximum logarithmic 
likelihood is $\log(L)=-3.41$, which is enclosed by 
the inner surface (blue) denoting $\log(L)=-4$.  The outer 
surface (red) has $\log(L)=-9.9$. If the likelihood function 
were Gaussian, the outer surface would be approximately at 
the ``1-$\sigma$'' level. 
The contour lines on the planes are linearly spaced in logarithm
with a spacing of 13.4. 
The figure shows that the parameters are strongly correlated in the 
plotted range and that the likelihood function 
changes dramatically with $f_{\rm SF}$. Within a small range of $\log f_{\rm SF}$, 
from 1.8 to 2.0, the iso-likelihood surface moves a long distance 
in the $\alpha_{\rm SF}$-$\alpha_{\rm RH}$ plane and becomes elongated 
when $\log f_{\rm SF}$ is close to 2.0.  
This type of complex structure and dramatic change happens also in 
the other dimensions. As a consequence, the posterior
is significantly more complex than one might expect by only looking at the
fully marginalised distributions presented in Figure \ref{fig:par_mod0}.  
The fine structure and complex topology of the posterior also make it clear
that it is extremely difficult to find the best fit by tuning model
parameters by hand.  It also explains why it is extremely computationally
challenging to properly sample the posterior. The MCMC algorithm must
navigate along extremely thin and curved surfaces in multiple dimensions,
often with very small gradients in likelihood to find the most probable models.
For example the results presented here required more than a million
SAM evaluations with approximately $5\times 10^4$ 2GHz Opteron CPU hours.

\section{The Impact of Prior Choice}
\label{sec:res_pri}

\begin{figure}
  \centering
  \includegraphics[width=\textwidth]{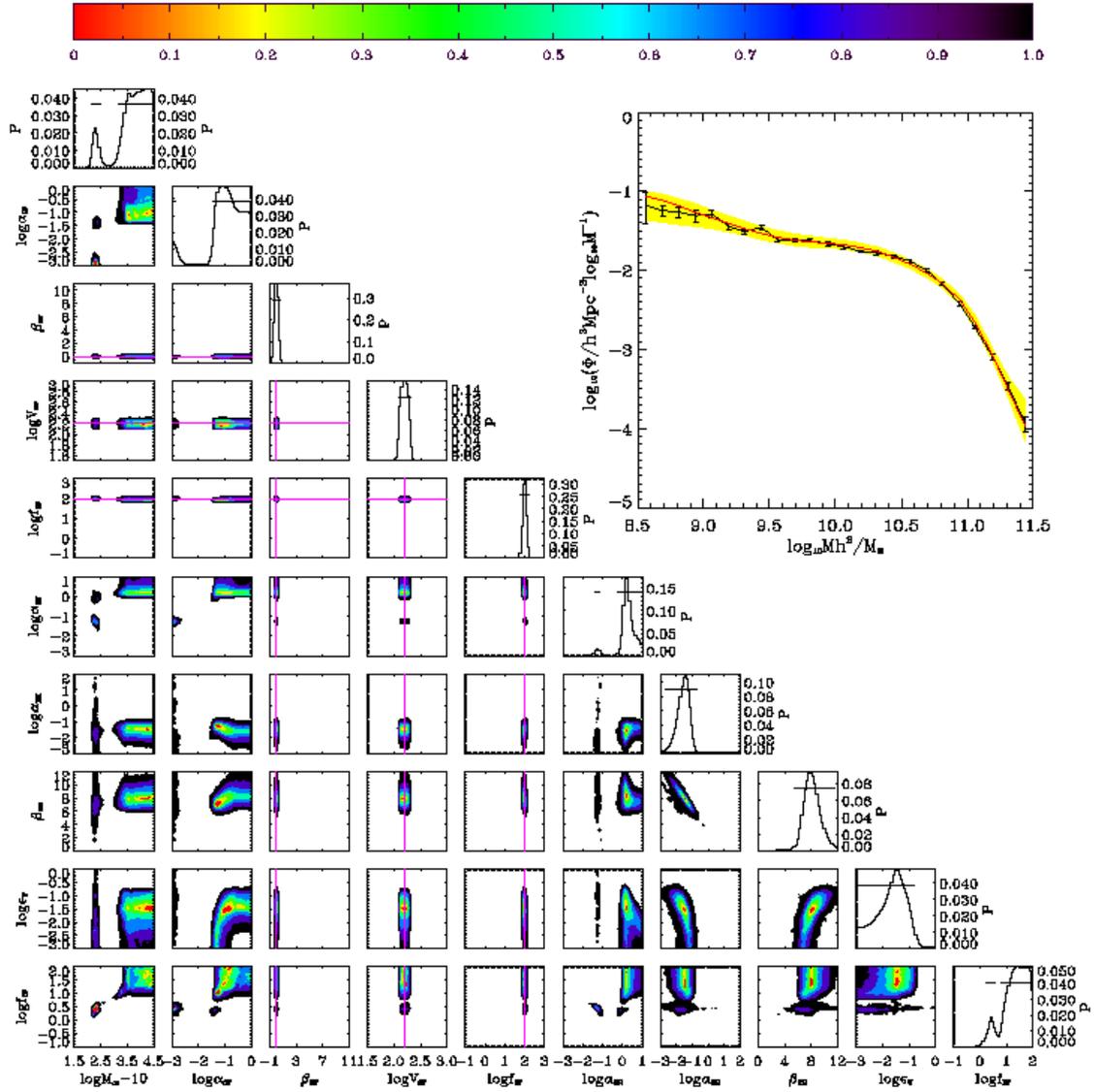}
  \caption {The marginalised posterior distribution for key parameters
    in Case 1.Very restrictive priors are assumed for the parameters,
    $\beta_{\rm SF}$, $V_{\rm SF}$ and $f_{\rm SF}$, whose central
    values are indicated by magenta lines.}
  \label{fig:par_mod1}
\end{figure}

In this section, we study the affect of the prior distribution on the
final inference by selectively applying narrow prior distributions for some of
the parameters.  This mimics the common practise of fixing some model
parameters.  In Case 1, three of the 13 parameters are given narrow
priors. The value of $\beta_{\rm SF}$ is limited to the narrow range
$[-0.2, 0.2]$, to mimic a flat power-law for star formation efficiency
as adopted in some SAMs \citep[e.g.][]{Croton2006}.  The
parameter $V_{\rm SF}$ has little effect so we set $\log (V_{\rm
  SF}/{\rm km\,s^{-1}})$ in the narrow range $[2.1, 2.3]$.
Furthermore, we assign a narrow prior, $[1.9, 2.1]$, to $\log f_{\rm
  SF}$, corresponding to the choice $\Sigma_{\rm sf}\approx
10\msun/{\rm pc}^2$ and $r_{\rm disc}\approx 3 r_{\rm disc0}$, which is
often used in previous SAMs \citep[e.g.][]{Croton2006}.  All the other
prior distributions are the same as in the fiducial case considered
\S\ref{sec:res_pos} (Case 0).  The resulting marginalised posterior
distributions in Figure \ref{fig:par_mod1}, and summarised in 
Table \ref{tab:par} as the second prior and posterior 95\% confidence
range listed for each parameter, 
show that the distribution
of all the parameters becomes more compact.  The improvement of prior
information on some parameters not only tightens the constraints on
those parameters themselves but can also help break degeneracies in
other dimensions.  For example, since the star formation law is
restricted to have weak dependence on halo mass, the efficiency of SN
reheating is forced to be a steep power-law of halo circular velocity.
For similar reasons, the degeneracies of the other parameters with 
$\beta_{\rm RH}$ are all also reduced.  Note that this restrictive prior is not
located near the dominant posterior mode with unrestricted priors
(cf. Fig. \ref{fig:par_mod0}).  Moreover, the bulk of the Case 1
posterior has very low probability in the Case 0 posterior.
Nevertheless, the quality of the fit does not change much, as one can
see from the reproduced stellar mass function shown in the
upper-right panel of Figure \ref{fig:par_mod1}. This illustrates the danger in
fixing the values of parameters to plausible values especially when there is
no compelling prior reason for imposing such a strong constraint.

\begin{figure}
  \centering
  \includegraphics[width=\textwidth]{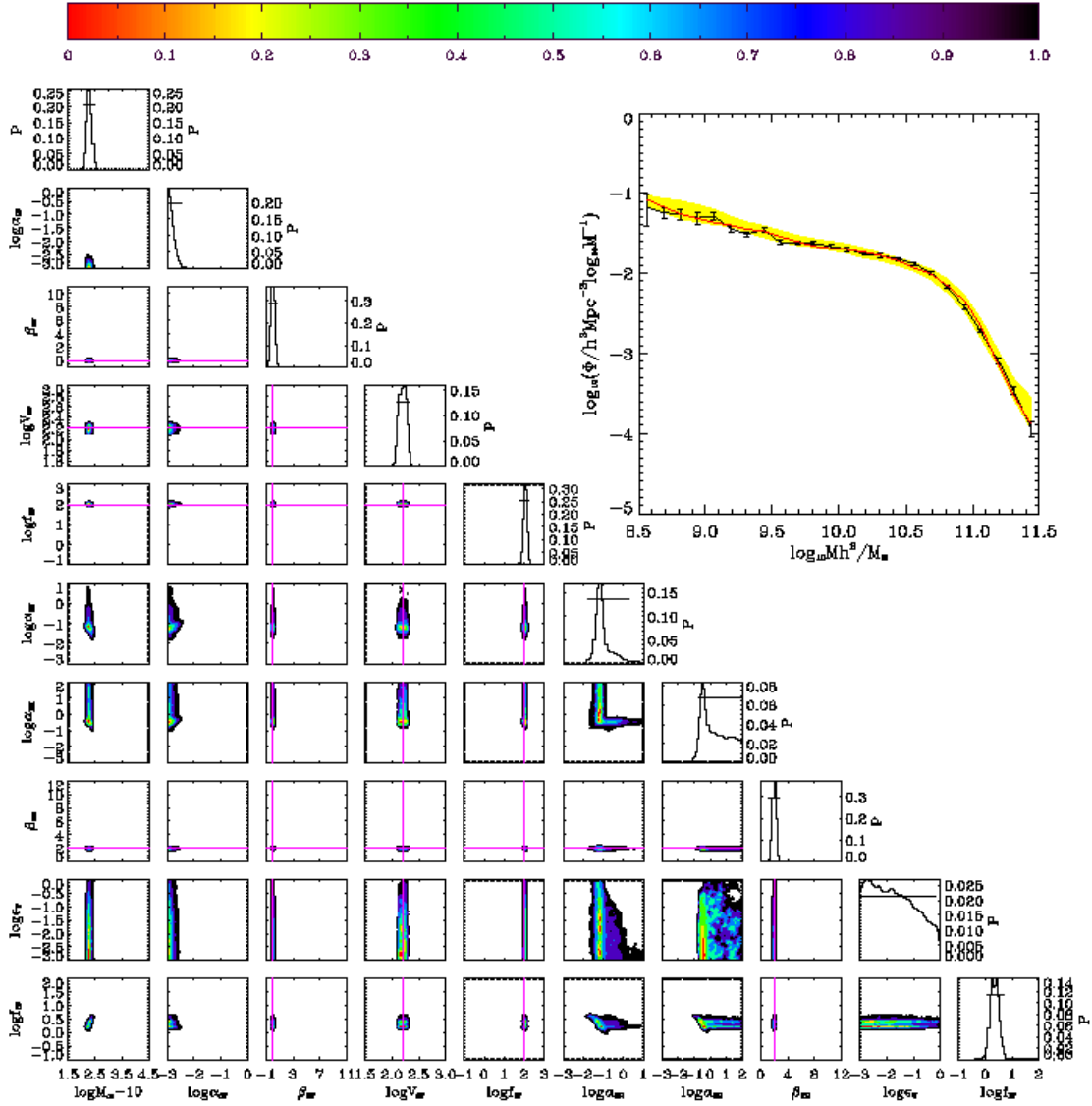}
  \caption {The marginalised posterior distribution of key parameters
    for Case 2. This includes the restrictions of Case 1 as in
    Fig. \protect{\ref{fig:par_mod1}} with an additional restrictive
    prior for $\beta_{\rm RH}$.  }
  \label{fig:par_mod2}
\end{figure}

Case 1 requires that the SN feedback be a very steep function of the
halo circular velocity when the star formation efficiency is forced to
change slowly with halo mass.  Early SAMs 
\citep[e.g.][]{Kauffmann1999, Kang2005}
assumed a weak dependence of the SN feedback on halo
mass ($\beta_{\rm RH}\sim 2$) and concluded that the number of faint
galaxies are over-predicted if $\beta_{\rm SF}\sim 0$.  However,
whether or not a good fit can still be obtained by changing the other
parameters while keeping $\beta_{\rm SF}\sim 0$ and $\beta_{\rm
  RH}\sim 2$ requires a full exploration of the high-dimensional
parameter space.
Case 2 addresses this question by imposing the
additional prior restriction, $\beta_{\rm RH} \in [1.8,2.2]$, 
and in Figure \ref{fig:par_mod2} we show 
the resulting marginalised distributions, which are summarised in the third
entry for each parameter in Table \ref{tab:par}.
The modes have moved substantially with respect to those in Case 1. To
compensate for the weakened SN reheating in small halos owing to the
assumed weak dependence of SN reheating on halo circular velocity, the
model employs a much lower efficiency for star formation and a larger
reheating amplitude; the mode moves from the lower-right to the
upper-left in the $\log \alpha_{\rm SF}-\log \alpha_{\rm RH}$ plane.
For similar reasons, the posterior mode in the other dimensions also
change.

The posterior-weighted stellar mass function 
shown in the upper-right panel of Figure \ref{fig:par_mod2}
still reproduces the observed stellar mass function
even though the power indices $\beta_{\rm SF}$
and $\beta_{\rm RH}$ are both fixed to low values.  This illustrates the
importance of carefully specifying the prior distribution for each
parameter, especially when a parameter is weakly constrained, and the
necessity for fully characterising the posterior distribution over its
full domain.

In summary, the results obtained in this section show that assigning
restrictive prior distributions to some parameters can significantly
reduce the volume of the parameter space and tighten the constraints
on all the parameters in SAMs of galaxy formation. Thus, 
any prior knowledge, either from observation or physical consideration, 
can help the model inference and hence improve our understanding of 
galaxy formation.  However, this also indicates that full posterior 
distribution can be very different from the posterior 
distribution with restricted priors
so that adopting unsubstantiated priors to fix parameters 
can lead to an erroneous inference for all the other parameters.
Also, it would not be straightforward to estimate the potential 
effects of introducing additional processes when the existing model parameters
are held fixed.
Because of the covariance between 
model parameters, adding a new model parameter may change the 
posterior significantly.

%%%%%%%%%%%%%%%%%%%%%%%%%%%%%%%%%%%%%%%%%%%%%%%%%%%%%%%%%%%%%%%%%%%%%%%%%%
\section{Impact of the Error Model}
\label{sec:res_cov}

The observational error model directly influences the value of
likelihood and the shape of the cost function in parameter
space. However, the impact of the error model has not been carefully
considered in SAMs. In this section, we explore the effect of
incorrect error estimates on the resulting inference.

Astronomers traditionally differentiate two kinds of errors,
\emph{statistical errors} and \emph{systematic errors}.  Statistical
errors result from well-understood processes with known parent
distributions (e.g. sampling error) while systematic errors come from
the underlying assumptions made for the measurements.  From the
Bayesian context, a \emph{systematic error} is the result of poor
prior information and often results in bias.  For the stellar mass
function considered here, the total error budget consists of the
independent statistical errors of individual stellar mass bins owing to
the finite number of galaxies used in estimating the stellar mass
function, and systematic errors, which arise from uncertainties in
the stellar population synthetic models used to estimate the stellar
mass from the observed luminosity. These systematic uncertainties correlate the
bins. For example, the uncertainty in the assumed 
IMF will increase or decrease the stellar masses of all the
galaxies in the same sense.

When errors in different mass bins are correlated, the likelihood
function may be approximated as follows:
\begin{equation}
  L(\mathbf{\Phi}_{\rm obs}| \mathbf{\theta}) 
= {L_0 \exp[-{1 \over 2} (\mathbf{\Phi}_{\rm obs} 
- \mathbf{\Phi}_{\rm mod})^T \cdot
\mathbf{\Sigma^{-1}} \cdot (\mathbf{\Phi}_{\rm obs} - \mathbf{\Phi}_{\rm mod})] 
\over (2 \pi)^{I/2} |{\rm det}(\mathbf{\Sigma})|^{1/2}},
\label{eq:Lcov}
\end{equation}
where $\mathbf{\Phi}_{\rm obs}$ and $\mathbf{\Phi}_{\rm mod}$ are the
vectors of the observed and predicted stellar mass functions over the
stellar mass bins, $\Sigma$ is the covariance matrix that describes
the correlated error model, and $I$ is the rank of the covariance
matrix. For independent errors, all the off-diagonal terms vanish and
the likelihood reduces to Eq.(\ref{eq:lik_ind}).

To test the effect of correlated error, we construct a synthetic
stellar mass function with correlated errors that mimic the systematic
uncertainties in real observation.  We choose a truncated series of
Chebyshev polynomials to represent the observed stellar mass
function. The low-order coefficients specify the overall shape of the
function, while the higher orders characterise small scale features.
We find that Chebyshev polynomials up to order 4 nicely fits the
logarithmic stellar mass function, $\log \Phi(\log m_*)$; the best-fit
coefficients are $[-4.17, -1.26,-0.516, -0.274, -0.114]$.  We choose
the standard deviations of these coefficients $[0.05, 0.10, 0.12,
0.08, 0.03]$ to represent the correlated uncertainties in the
measurements.  Then, we calculate the covariance matrix of this
synthetic data using 1000 Monte Carlo realisations of the Chebyshev
polynomials and use the synthetic data and the derived covariance matrix
to constrain the parameters in our SAM.

\begin{figure}
    \hfill
    \begin{minipage}[t]{1.00\textwidth}
    \begin{center}
        \epsfig{file=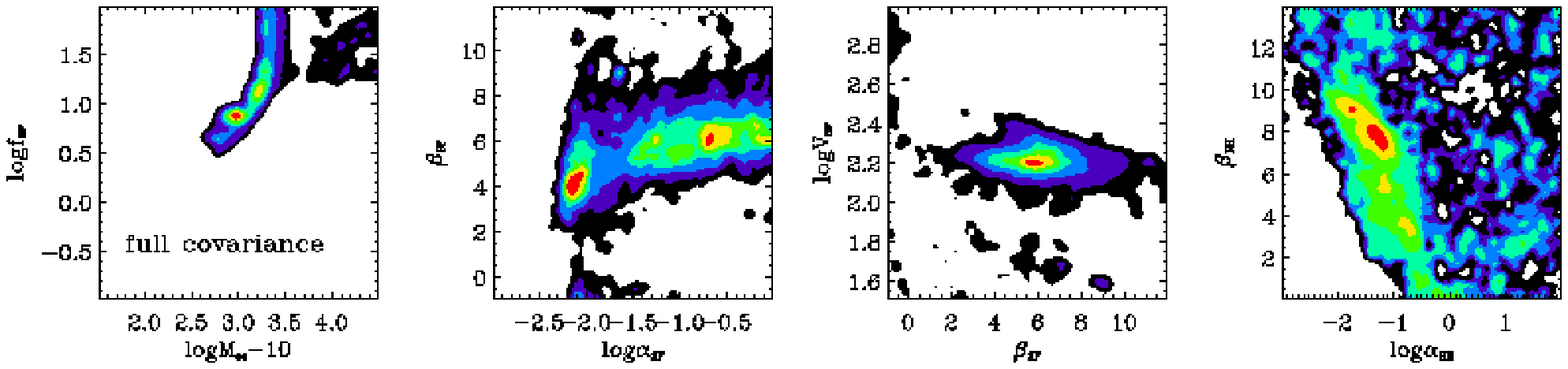, scale=0.7}
    \end{center}
    \end{minipage}
    \vfill

    \hfill
    \begin{minipage}[t]{1.00\textwidth}
        \begin{center}
            \epsfig{file=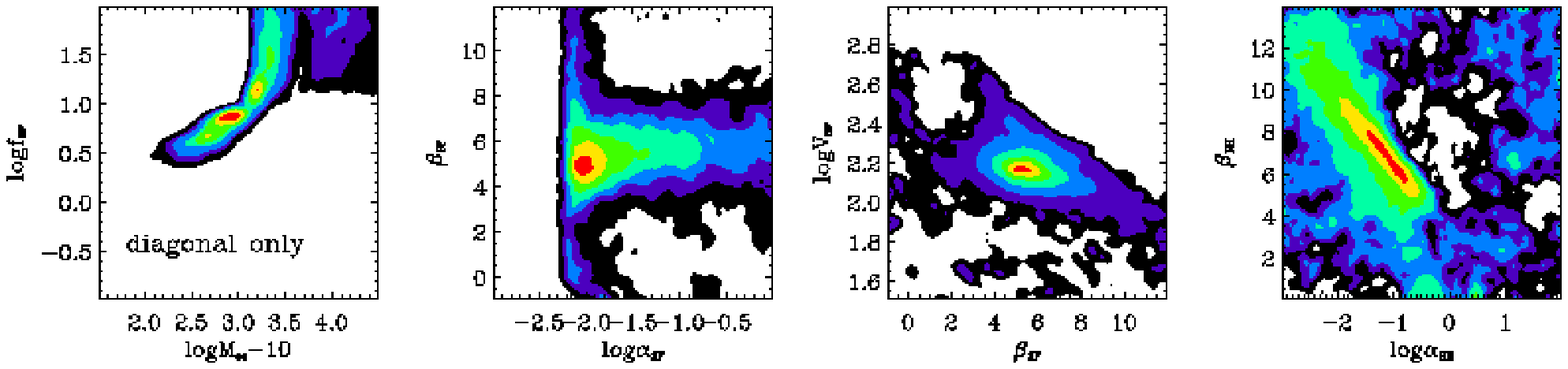, scale=0.7}
        \end{center}
    \end{minipage}
    \caption{A comparison of the posterior distribution obtained for a
      likelihood function including covariance (upper row,
      eq. \protect{\ref{eq:Lcov}}), and using only the diagonal terms
      of the covariance matrix (lower row).}\label{fig:cov}
\end{figure}

Figure \ref{fig:cov} compares the marginalised posteriori for four
pairs of model parameters obtained with the full covariance matrix
(upper panels) and those obtained with the diagonal terms only (lower
panels).  Removing the off-diagonal terms is equivalent to ignoring
the covariance. Clearly, the contours produced with the full covariance
matrix are more compact. This is expected because the correlation of the
errors between the different bins implies that the total independent error
in the data is smaller.  There are also noticeable changes in the
shape and the orientation of the posterior distribution, indicating
that it is important to model the covariance of the data properly
to make correct model inferences. For example, in the 
$\beta_{\rm SF}$-$\log\alpha_{\rm SF}$ plane a new mode appears with
$\log\alpha_{\rm SF}\sim 0.5$.

%%%%%%%%%%%%%%%%%%%%%%%%%%%%%%%%%%%%%%%%%%%%%%%%%%%%%%%%%%%%%%%%%%%%%%%%%%
\section{Summary and Discussion}
\label{sec:dis}

Many of the physical processes parametrised in semi-analytical models
of galaxy formation remain poorly understood and under specified.
This has two critically important consequences for inferring
constraints on the physical parameters: 1) prior assumptions about the
size of the domain and the shape of the parameter distribution will
strongly affect any resulting inference; and 2) a very large parameter
space must be fully explored to obtain an accurate inference.
Moreover, both \emph{must} be done together.  Both of these issues are
naturally tackled with a Bayesian approach that allows one to
constrain theory with data in a probabilistically rigorous way.  In
this paper, we have presented a semi-analytic model of galaxy
formation in the framework of Bayesian inference and illustrated its
performance on a test problem.  Our sixteen-parameter semi-analytic
model incorporates the most commonly used parameterisations of
important physical processes from existing SAMs including star
formation, SN feedback, galaxy merger, and AGN feedback.  We combined
this model with the Bayesian Inference Engine developed at
the University of Massachusetts.
The BIE is an extensible MPI-based software package for
developing, testing and running advanced Markov-Chain Monte-Carlo
algorithms on large data sets.  The resulting tool allows the
exploration of the posterior distribution of a large number of model
parameters, and to constrain models over multiple data sets in a
statistically rigorous way.

To demonstrate the power of this approach, we used the observationally
derived stellar mass function of galaxies to constrain a number of
important model parameters and to study the posterior probability 
distribution of the model parameters. 
The posterior probability is the conditional probability
obtained after the data is taken into account.
This is an important quantity for model inference because it 
specifies, for a given model family,  the probability distribution 
of the parameters in the full parameter space. 
The Bayesian model inference requires one to examine the full 
posterior probability distribution, emphasising the 
marginalised probability instead of the `best' parameters. 
We find that the posterior distribution
has a very complex structure and topology, indicating that finding the
best fit by tweaking model parameters is improbable.  
Moreover, the posterior clearly shows that
many model parameters are strongly covariant and, therefore, the
inferred value of a particular parameter can be significantly affected
by the priors used for the other parameters.  
As a consequence, one needs to have the knowledge about the posterior distribution 
in the entire parameter space when making a model inference, because it is very likely 
to miss significant modes if some parameters are fixed.
We have demonstrated here that restricting the prior of some parameters without physical reasons
can exclude models that should be considered as ``plausible'', and hence
lead to biased inferences. We note that the ``implausibility'' proposed by \citet{Bower2010} 
is a measure of the adequacy of a fit defined 
for individual parameter sets. This is different from the posterior distribution discussed here.
Finally, with the use of synthetic data to mimic
systematic uncertainties in the reduced data, we have shown that
model parameter inferences can be significantly affected by
the use of an incorrect error model.  This clearly demonstrates that an
accurate error model (both sampling error and systematic
uncertainties) is crucial to using observational data correctly, and
conversely, a data-model comparison without an accurate error model is
likely to be erroneous.

The method developed here can be straightforwardly applied to other
data sets and to multiple data sets simultaneously.  Large galaxy
surveys available now and in the near future will provide many more
data sets to characterise the properties of the galaxy population not
only at the present time but also at high redshifts.  The
Bayesian-inference based SAM described in this paper provides a
framework for parameter estimation (e.g constraining the parameters
in theoretical models given particular data sets), for hypothesis testing
(e.g. comparing the support for particular models given the data), and
for predicting the power of new observations to constraining models of
interest.  In addition, the Bayesian approach explicitly builds on
previous results by incorporating the constraints from previous
inferences into new data sets; the Bayesian Inference Engine is
designed to do this automatically.  The approach developed here will
produce probabilistically rigorous constraints on theoretical models,
and facilitate understanding the underlying physical processes that shape
the observed galaxy population.  For many processes in galaxy
formation, competing models have been proposed but not quantitatively
compared.  The marginalised likelihood or \emph{Bayes Evidence}, which 
can be directly derived from the posterior, and explicit model comparison
techniques, such as the reversible jump algorithm \citep{Green1995},
can provide a quantitative comparison between different models for given data.
The Bayesian hypothesis test can, therefore, be used to discriminate between
models.  Finally, the prediction for an observable including the
inferential uncertainties can be obtained by marginalising over the posterior.
Such predictions can be used to assess the power of new observations.
In a series of forthcoming papers, we will use the scheme developed
here to make inferences from various data sets, focusing on a number
of the aspects discussed above.

It should be pointed out, however, that the method developed here 
has its limitations, and improvements in the methodology are still 
needed to realise its full potential. For example, since 
the posterior is explored using MCMC sampling, one expects the computation  
to be more challenging as the dimensionality of the
problem increases. Higher dimensionality means a larger parameter 
volume to be explored. If the new dimensions do not have a strong
impact on the model predictions, the posterior will be extended in 
the new dimensions and require longer chains and/or more parallel 
chains to sample. If, on the other hand, the posterior is very compact
in the new dimensions, the chance for a MCMC chain to find a mode 
will decrease. In practise, we expect the situation to be between these 
two extremes, but both cases make the computation more challenging. 
In addition, as the complexity of the model increases, one expects 
the topology of the posterior distribution to become more complicated, 
making it harder for the MCMC chains to mix well. We note that these 
general difficulties challenge any method of parameter space exploration, 
so one has to try different algorithms to determine 
the one that is the best suited for the problem in question.
For the Tempered Differential Evolution MCMC algorithm adopted
here, we may have to increase the maximum temperature for tempering
to enhance mixing and/or increase the number of chains. 
Both of these will increase the number of SAM evaluations, each 
taking of order one minute to compute. One way to increase 
the speed of the 
likelihood evaluation would be to adopt a model emulator, such 
as the one proposed by \citet{Bower2010}. 
However, we note that significant 
computation is required to train such an emulator, and that the 
error introduced by the emulator has to be carefully controlled.

The adding of new data to constrain the model will alter the posterior
distribution. For example, the additional data may reduce the
credible volume and/or produce more complex features in the
posterior distribution.  Except for increasing the computation time
and/or improving the sampling efficiency of the algorithm as described
above, the Bayesian update \citep[e.g.][]{Weinberg2010} provides a
natural solution to the problem.  The data-based hierarchical
approach, eliciting prior information from the posterior distribution
constrained by less data, allows us to update the posterior as more
and more data are included into the inference.  This approach may
reduce the computation time significantly.  On the other hand,
however, it is also possible that the additional data is not mutually
compatible with the old data and the main modes would change
significantly. Finally, the compatibility of the model with the two
data sets, e.g. how well the model accommodates the data, should be
assessed with a goodness-of-fit test.

Although the Bayesian approach is conceptually attractive, these
computations are very costly.  One way to increase the speed of the
likelihood evaluation would be to adopt a model emulator, such as the
one proposed by \citet{Bower2010}.  However, we note that the error
introduced by the emulator has to be carefully controlled.  For
example, Figure \ref{fig:pos_cut} illustrates the complexity of the
posterior distribution in parameter space.  It remains to be verified
that such a correlation can be accurately represented by an emulator.
If the posterior distribution from the emulator approach could be used
in some way to provide a prior distribution for the direct simulation, the
two methods together may prove to be more powerful than either alone.
\citet{Bower2010} proposed an implausibility measure to
characterise the fit of the predictions to the data.  Like the prior
distribution, the implausibility reduces the parameter space to be
explored before new data is included.  This refining inference can
also be accomplished by a Bayesian update, which uses the posterior
obtained from part of the data as the prior for the next level of the
inference that uses the remaining data.

As mentioned earlier, any model can only be considered as an
approximation to reality, because of model uncertainties including
stochasticity of the involved processes and inaccuracy of the model
prescriptions. These uncertainties may be included in the likelihood
if an assumption is made about their statistical properties. The
`model discrepancy' introduced in \citet{Bower2010} is an attempt
along these lines. A large amount of the model uncertainties arise in
SAMs because some processes are not well understood and can be
modelled in various plausible ways. If a specific set of prescriptions
is adopted by the model without taking into account their
uncertainties, any inferences made are only for the model family
specified by the adopted prescriptions. To include the model
uncertainties, one can generalise the prescriptions to encompass a
large number of possibilities. Alternatively, one can consider a
number of plausible model families and use Bayes Evidence to
discriminate between the different model families. In a forthcoming
paper, we will demonstrate how the marginalised likelihood or
\emph{Bayes Evidence} can provide a quantitative comparison between
different models for given data.

%%%%%%%%%%%%%%%%%%%%%%%%%%%%%%%%%%%%%%%%%%%%%%%%%%%%%%%%%%%%%%%%%%%%%%%%%%
% Acknowledgements
%%%%%%%%%%%%%%%%%%%%%%%%%%%%%%%%%%%%%%%%%%%%%%%%%%%%%%%%%%%%%%%%%%%%%%%%%%
\section*{Acknowledgements}
This material is based upon work supported by the National Aeronautics
and Space Administration AISR program AISR-126270 and the National
Science Foundation under Grant No. III-0611948.  Disclaimer: Any
opinions, findings, and conclusions or recommendations expressed in
this material are those of the authors and do not necessarily reflect
the views of the National Science Foundation.

%%%%%%%%%%%%%%%
% Bibliography
%%%%%%%%%%%%%%%%%%%%%%%%%%%%%%%%%%%%%%%%%%%%%%%%%%%%%%%%%%%%%%%%%%%%%%%%%%

\bibliography{/Users/luyu/references/general}

\begin{thebibliography}{83}
\expandafter\ifx\csname natexlab\endcsname\relax\def\natexlab#1{#1}\fi

\bibitem[{{Avila-Reese} \& {Firmani}(2000)}]{Avila-Reese2000}
{Avila-Reese} V., {Firmani} C., 2000, \rmxaa, 36, 23

\bibitem[{{Avila-Reese} {et~al.}(1998){Avila-Reese}, {Firmani}, \&
  {Hern{\'a}ndez}}]{Avila-Reese1998}
{Avila-Reese} V., {Firmani} C., {Hern{\'a}ndez} X., 1998, \apj, 505, 37

\bibitem[{{Baldry} {et~al.}(2006){Baldry}, {Balogh}, {Bower}, {Glazebrook},
  {Nichol}, {Bamford}, \& {Budavari}}]{Baldry2006}
{Baldry} I.~K., {Balogh} M.~L., {Bower} R.~G., {Glazebrook} K., {Nichol} R.~C.,
  {Bamford} S.~P., {Budavari} T., 2006, \mnras, 373, 469

\bibitem[{{Bell} {et~al.}(2003){Bell}, {McIntosh}, {Katz}, \&
  {Weinberg}}]{Bell2003}
{Bell} E.~F., {McIntosh} D.~H., {Katz} N., {Weinberg} M.~D., 2003, \apjs, 149,
  289

\bibitem[{{Benson} \& {Bower}(2010)}]{Benson2010a}
{Benson} A.~J., {Bower} R., 2010, \mnras, 405, 1573

\bibitem[{{Benson} {et~al.}(2003){Benson}, {Bower}, {Frenk}, {Lacey}, {Baugh},
  \& {Cole}}]{Benson2003c}
{Benson} A.~J., {Bower} R.~G., {Frenk} C.~S., {Lacey} C.~G., {Baugh} C.~M.,
  {Cole} S., 2003, \apj, 599, 38

\bibitem[{{Benson} \& {Madau}(2003)}]{Benson2003}
{Benson} A.~J., {Madau} P., 2003, \mnras, 344, 835

\bibitem[{{Bertschinger}(1985)}]{Bertschinger1985}
{Bertschinger} E., 1985, \apjs, 58, 39

\bibitem[{{Bett} {et~al.}(2010){Bett}, {Eke}, {Frenk}, {Jenkins}, \&
  {Okamoto}}]{Bett2010}
{Bett} P., {Eke} V., {Frenk} C.~S., {Jenkins} A., {Okamoto} T., 2010, \mnras,
  256

\bibitem[{{Bigiel} {et~al.}(2008){Bigiel}, {Leroy}, {Walter}, {Brinks}, {de
  Blok}, {Madore}, \& {Thornley}}]{Bigiel2008}
{Bigiel} F., {Leroy} A., {Walter} F., {Brinks} E., {de Blok} W.~J.~G., {Madore}
  B., {Thornley} M.~D., 2008, \aj, 136, 2846

\bibitem[{{Binney} \& {Tremaine}(1987)}]{Binney1987}
{Binney} J., {Tremaine} S., 1987, {Galactic dynamics}. Princeton, NJ, Princeton
  University Press, 1987, 747 p.

\bibitem[{{Bond} {et~al.}(1991){Bond}, {Cole}, {Efstathiou}, \&
  {Kaiser}}]{Bond1991}
{Bond} J.~R., {Cole} S., {Efstathiou} G., {Kaiser} N., 1991, \apj, 379, 440

\bibitem[{{Bower} {et~al.}(2006){Bower}, {Benson}, {Malbon}, {Helly}, {Frenk},
  {Baugh}, {Cole}, \& {Lacey}}]{Bower2006}
{Bower} R.~G., {Benson} A.~J., {Malbon} R., {Helly} J.~C., {Frenk} C.~S.,
  {Baugh} C.~M., {Cole} S., {Lacey} C.~G., 2006, \mnras, 370, 645

\bibitem[{{Bower} {et~al.}(2010){Bower}, {Vernon}, {Goldstein}, {Benson},
  {Lacey}, {Baugh}, {Cole}, \& {Frenk}}]{Bower2010}
{Bower} R.~G., {Vernon} I., {Goldstein} M., {Benson} A.~J., {Lacey} C.~G.,
  {Baugh} C.~M., {Cole} S., {Frenk} C.~S., 2010, \mnras, 407, 2017

\bibitem[{{Bruzual} \& {Charlot}(2003)}]{Bruzual2003}
{Bruzual} G., {Charlot} S., 2003, \mnras, 344, 1000

\bibitem[{{Bullock} {et~al.}(2001{\natexlab{a}}){Bullock}, {Dekel}, {Kolatt},
  {Kravtsov}, {Klypin}, {Porciani}, \& {Primack}}]{Bullock2001}
{Bullock} J.~S., {Dekel} A., {Kolatt} T.~S., {Kravtsov} A.~V., {Klypin} A.~A.,
  {Porciani} C., {Primack} J.~R., 2001{\natexlab{a}}, \apj, 555, 240

\bibitem[{{Bullock} {et~al.}(2001{\natexlab{b}}){Bullock}, {Kolatt}, {Sigad},
  {Somerville}, {Kravtsov}, {Klypin}, {Primack}, \& {Dekel}}]{Bullock2001a}
{Bullock} J.~S., {Kolatt} T.~S., {Sigad} Y., {Somerville} R.~S., {Kravtsov}
  A.~V., {Klypin} A.~A., {Primack} J.~R., {Dekel} A., 2001{\natexlab{b}},
  \mnras, 321, 559

\bibitem[{{Cattaneo} {et~al.}(2006){Cattaneo}, {Dekel}, {Devriendt},
  {Guiderdoni}, \& {Blaizot}}]{Cattaneo2006}
{Cattaneo} A., {Dekel} A., {Devriendt} J., {Guiderdoni} B., {Blaizot} J., 2006,
  \mnras, 370, 1651

\bibitem[{{Cole} \& {Lacey}(1996)}]{Cole1996}
{Cole} S., {Lacey} C., 1996, \mnras, 281, 716

\bibitem[{{Cole} {et~al.}(2000){Cole}, {Lacey}, {Baugh}, \& {Frenk}}]{Cole2000}
{Cole} S., {Lacey} C.~G., {Baugh} C.~M., {Frenk} C.~S., 2000, \mnras, 319, 168

\bibitem[{{Croton} {et~al.}(2006){Croton}, {Springel}, {White}, {De Lucia},
  {Frenk}, {Gao}, {Jenkins}, {Kauffmann}, {Navarro}, \& {Yoshida}}]{Croton2006}
{Croton} D.~J., {Springel} V., {White} S.~D.~M., {De Lucia} G., {Frenk} C.~S.,
  {Gao} L., {Jenkins} A., {Kauffmann} G., {Navarro} J.~F., {Yoshida} N., 2006,
  \mnras, 365, 11

\bibitem[{{De Lucia} \& {Blaizot}(2007)}]{DeLucia2007}
{De Lucia} G., {Blaizot} J., 2007, \mnras, 375, 2

\bibitem[{{Dunkley} {et~al.}(2009){Dunkley}, {Komatsu}, {Nolta}, {Spergel},
  {Larson}, {Hinshaw}, {Page}, {Bennett}, {Gold}, {Jarosik}, {Weiland},
  {Halpern}, {Hill}, {Kogut}, {Limon}, {Meyer}, {Tucker}, {Wollack}, \&
  {Wright}}]{Dunkley2009}
{Dunkley} J., {Komatsu} E., {Nolta} M.~R., {Spergel} D.~N., {Larson} D.,
  {Hinshaw} G., {Page} L., {Bennett} C.~L., {Gold} B., {Jarosik} N., {Weiland}
  J.~L., {Halpern} M., {Hill} R.~S., {Kogut} A., {Limon} M., {Meyer} S.~S.,
  {Tucker} G.~S., {Wollack} E., {Wright} E.~L., 2009, \apjs, 180, 306

\bibitem[{{Dutton} \& {van den Bosch}(2009)}]{Dutton2009}
{Dutton} A.~A., {van den Bosch} F.~C., 2009, \mnras, 396, 141

\bibitem[{{Dutton} {et~al.}(2007){Dutton}, {van den Bosch}, {Dekel}, \&
  {Courteau}}]{Dutton2007}
{Dutton} A.~A., {van den Bosch} F.~C., {Dekel} A., {Courteau} S., 2007, \apj,
  654, 27

\bibitem[{{Efstathiou} {et~al.}(1985){Efstathiou}, {Davis}, {White}, \&
  {Frenk}}]{Efstathiou1985}
{Efstathiou} G., {Davis} M., {White} S.~D.~M., {Frenk} C.~S., 1985, \apjs, 57,
  241

\bibitem[{{Fardal} {et~al.}(2007){Fardal}, {Katz}, {Weinberg}, \&
  {Dav{\'e}}}]{Fardal2007}
{Fardal} M.~A., {Katz} N., {Weinberg} D.~H., {Dav{\'e}} R., 2007, \mnras, 379,
  985

\bibitem[{{Fillmore} \& {Goldreich}(1984)}]{Fillmore1984}
{Fillmore} J.~A., {Goldreich} P., 1984, \apj, 281, 1

\bibitem[{{Firmani} \& {Avila-Reese}(2000)}]{Firmani2000}
{Firmani} C., {Avila-Reese} V., 2000, \mnras, 315, 457

\bibitem[{{Fontanot} {et~al.}(2009){Fontanot}, {De Lucia}, {Monaco},
  {Somerville}, \& {Santini}}]{Fontanot2009}
{Fontanot} F., {De Lucia} G., {Monaco} P., {Somerville} R.~S., {Santini} P.,
  2009, \mnras, 397, 1776

\bibitem[{{Fu} {et~al.}(2010){Fu}, {Guo}, {Kauffmann}, \& {Krumholz}}]{Fu2010}
{Fu} J., {Guo} Q., {Kauffmann} G., {Krumholz} M.~R., 2010, ArXiv e-prints

\bibitem[{Gelman \& Rubin(1992)}]{Gelman1992}
Gelman A., Rubin D., 1992, Statistical Science, 7, 457

\bibitem[{Green(1995)}]{Green1995}
Green P.~J., 1995, Biometrika, 82, 711

\bibitem[{{Gunn} \& {Gott}(1972)}]{Gunn1972}
{Gunn} J.~E., {Gott} I. J.~R., 1972, \apj, 176, 1

\bibitem[{{Henriques} {et~al.}(2009){Henriques}, {Thomas}, {Oliver}, \&
  {Roseboom}}]{Henriques2009}
{Henriques} B.~M.~B., {Thomas} P.~A., {Oliver} S., {Roseboom} I., 2009, \mnras,
  396, 535

\bibitem[{{Kampakoglou} {et~al.}(2008){Kampakoglou}, {Trotta}, \&
  {Silk}}]{Kampakoglou2008}
{Kampakoglou} M., {Trotta} R., {Silk} J., 2008, \mnras, 384, 1414

\bibitem[{{Kang} {et~al.}(2005){Kang}, {Jing}, {Mo}, \&
  {B{\"o}rner}}]{Kang2005}
{Kang} X., {Jing} Y.~P., {Mo} H.~J., {B{\"o}rner} G., 2005, \apj, 631, 21

\bibitem[{{Katz}(1992)}]{Katz1992}
{Katz} N., 1992, \apj, 391, 502

\bibitem[{{Kauffmann} {et~al.}(1999){Kauffmann}, {Colberg}, {Diaferio}, \&
  {White}}]{Kauffmann1999}
{Kauffmann} G., {Colberg} J.~M., {Diaferio} A., {White} S.~D.~M., 1999, \mnras,
  303, 188

\bibitem[{{Kauffmann} {et~al.}(1993){Kauffmann}, {White}, \&
  {Guiderdoni}}]{Kauffmann1993}
{Kauffmann} G., {White} S.~D.~M., {Guiderdoni} B., 1993, \mnras, 264, 201

\bibitem[{{Kennicutt}(1998)}]{Kennicutt1998}
{Kennicutt} J. R.~C., 1998, \araa, 36, 189

\bibitem[{{Kennicutt} {et~al.}(2007){Kennicutt}, {Calzetti}, {Walter}, {Helou},
  {Hollenbach}, {Armus}, {Bendo}, {Dale}, {Draine}, {Engelbracht}, {Gordon},
  {Prescott}, {Regan}, {Thornley}, {Bot}, {Brinks}, {de Blok}, {de Mello},
  {Meyer}, {Moustakas}, {Murphy}, {Sheth}, \& {Smith}}]{Kennicutt2007}
{Kennicutt} J. R.~C., {Calzetti} D., {Walter} F., {Helou} G., {Hollenbach}
  D.~J., {Armus} L., {Bendo} G., {Dale} D.~A., {Draine} B.~T., {Engelbracht}
  C.~W., {Gordon} K.~D., {Prescott} M.~K.~M., {Regan} M.~W., {Thornley} M.~D.,
  {Bot} C., {Brinks} E., {de Blok} E., {de Mello} D., {Meyer} M., {Moustakas}
  J., {Murphy} E.~J., {Sheth} K., {Smith} J.~D.~T., 2007, \apj, 671, 333

\bibitem[{{Kere\v{s}} {et~al.}(2005){Kere\v{s}}, {Katz}, {Weinberg}, \&
  {Dav{\'e}}}]{Kerevs2005}
{Kere\v{s}} D., {Katz} N., {Weinberg} D.~H., {Dav{\'e}} R., 2005, \mnras, 363,
  2

\bibitem[{{Kimm} {et~al.}(2009){Kimm}, {Somerville}, {Yi}, {van den Bosch},
  {Salim}, {Fontanot}, {Monaco}, {Mo}, {Pasquali}, {Rich}, \&
  {Yang}}]{Kimm2009}
{Kimm} T., {Somerville} R.~S., {Yi} S.~K., {van den Bosch} F.~C., {Salim} S.,
  {Fontanot} F., {Monaco} P., {Mo} H., {Pasquali} A., {Rich} R.~M., {Yang} X.,
  2009, \mnras, 394, 1131

\bibitem[{{Komatsu} {et~al.}(2009){Komatsu}, {Dunkley}, {Nolta}, {Bennett},
  {Gold}, {Hinshaw}, {Jarosik}, {Larson}, {Limon}, {Page}, {Spergel},
  {Halpern}, {Hill}, {Kogut}, {Meyer}, {Tucker}, {Weiland}, {Wollack}, \&
  {Wright}}]{Komatsu2009}
{Komatsu} E., {Dunkley} J., {Nolta} M.~R., {Bennett} C.~L., {Gold} B.,
  {Hinshaw} G., {Jarosik} N., {Larson} D., {Limon} M., {Page} L., {Spergel}
  D.~N., {Halpern} M., {Hill} R.~S., {Kogut} A., {Meyer} S.~S., {Tucker} G.~S.,
  {Weiland} J.~L., {Wollack} E., {Wright} E.~L., 2009, \apjs, 180, 330

\bibitem[{{Krumholz} {et~al.}(2009){Krumholz}, {McKee}, \&
  {Tumlinson}}]{Krumholz2009}
{Krumholz} M.~R., {McKee} C.~F., {Tumlinson} J., 2009, \apj, 699, 850

\bibitem[{{Lacey} \& {Cole}(1993)}]{Lacey1993}
{Lacey} C., {Cole} S., 1993, \mnras, 262, 627

\bibitem[{{Liu} {et~al.}(2010){Liu}, {Yang}, {Mo}, {van den Bosch}, \&
  {Springel}}]{Liu2010}
{Liu} L., {Yang} X., {Mo} H.~J., {van den Bosch} F.~C., {Springel} V., 2010,
  \apj, 712, 734

\bibitem[{{Lu} {et~al.}(2010){Lu}, {Kere{\v s}}, {Katz}, {Mo}, {Fardal}, \&
  {Weinberg}}]{Lu2010}
{Lu} Y., {Kere{\v s}} D., {Katz} N., {Mo} H.~J., {Fardal} M., {Weinberg} M.~D.,
  2010, ArXiv e-prints

\bibitem[{{Lu} {et~al.}(2006){Lu}, {Mo}, {Katz}, \& {Weinberg}}]{Lu2006}
{Lu} Y., {Mo} H.~J., {Katz} N., {Weinberg} M.~D., 2006, \mnras, 368, 1931

\bibitem[{{Macci{\`o}} {et~al.}(2007){Macci{\`o}}, {Dutton}, {van den Bosch},
  {Moore}, {Potter}, \& {Stadel}}]{Maccio2007}
{Macci{\`o}} A.~V., {Dutton} A.~A., {van den Bosch} F.~C., {Moore} B., {Potter}
  D., {Stadel} J., 2007, \mnras, 378, 55

\bibitem[{{Martin} \& {Kennicutt}(2001)}]{Martin2001}
{Martin} C.~L., {Kennicutt} J. R.~C., 2001, \apj, 555, 301

\bibitem[{{Mckay} {et~al.}(1979){Mckay}, {Beckman}, \& {Conover}}]{Mckay1979}
{Mckay} M.~D., {Beckman} R.~J., {Conover} W.~J., 1979, Technometrics, 21, 239,
  american Statistical Association and American Society for Quality

\bibitem[{{Mo} {et~al.}(1998){Mo}, {Mao}, \& {White}}]{Mo1998}
{Mo} H.~J., {Mao} S., {White} S.~D.~M., 1998, \mnras, 295, 319

\bibitem[{{Mo} {et~al.}(2005){Mo}, {Yang}, {van den Bosch}, \& {Katz}}]{Mo2005}
{Mo} H.~J., {Yang} X., {van den Bosch} F.~C., {Katz} N., 2005, \mnras, 363,
  1155

\bibitem[{{Navarro} \& {Benz}(1991)}]{Navarro1991}
{Navarro} J.~F., {Benz} W., 1991, \apj, 380, 320

\bibitem[{{Navarro} {et~al.}(1997){Navarro}, {Frenk}, \& {White}}]{Navarro1997}
{Navarro} J.~F., {Frenk} C.~S., {White} S.~D.~M., 1997, \apj, 490, 493

\bibitem[{{Navarro} \& {White}(1993)}]{Navarro1993}
{Navarro} J.~F., {White} S.~D.~M., 1993, \mnras, 265, 271

\bibitem[{Neal(1996)}]{Neal1996}
Neal R., 1996, Statistics and Computing, 6, 353

\bibitem[{{Neistein} \& {Weinmann}(2010)}]{Neistein2010}
{Neistein} E., {Weinmann} S.~M., 2010, \mnras, 405, 2717

\bibitem[{{Oppenheimer} \& {Dav{\'e}}(2006)}]{Oppenheimer2006}
{Oppenheimer} B.~D., {Dav{\'e}} R., 2006, \mnras, 373, 1265

\bibitem[{{Parkinson} {et~al.}(2008){Parkinson}, {Cole}, \&
  {Helly}}]{Parkinson2008}
{Parkinson} H., {Cole} S., {Helly} J., 2008, \mnras, 383, 557

\bibitem[{{Press} \& {Schechter}(1974)}]{Press1974}
{Press} W.~H., {Schechter} P., 1974, \apj, 187, 425

\bibitem[{{Primack}(2009)}]{Primack2009}
{Primack} J.~R., 2009, in American Institute of Physics Conference Series, Vol.
  1192, American Institute of Physics Conference Series, {F.~Roig, D.~Lopes,
  R.~de La Reza, \& V.~Ortega}, ed., pp. 101--137

\bibitem[{{Primack} {et~al.}(2008){Primack}, {Gilmore}, \&
  {Somerville}}]{Primack2008}
{Primack} J.~R., {Gilmore} R.~C., {Somerville} R.~S., 2008, in American
  Institute of Physics Conference Series, Vol. 1085, American Institute of
  Physics Conference Series, {F.~A.~Aharonian, W.~Hofmann, \& F.~Rieger}, ed.,
  pp. 71--82

\bibitem[{{Sheth} {et~al.}(2001){Sheth}, {Mo}, \& {Tormen}}]{Sheth2001}
{Sheth} R.~K., {Mo} H.~J., {Tormen} G., 2001, \mnras, 323, 1

\bibitem[{{Simha} {et~al.}(2009){Simha}, {Weinberg}, {Dav{\'e}}, {Gnedin},
  {Katz}, \& {Kere\v{s}}}]{Simha2009}
{Simha} V., {Weinberg} D.~H., {Dav{\'e}} R., {Gnedin} O.~Y., {Katz} N.,
  {Kere\v{s}} D., 2009, \mnras, 399, 650

\bibitem[{{Somerville} {et~al.}(2008){Somerville}, {Hopkins}, {Cox},
  {Robertson}, \& {Hernquist}}]{Somerville2008}
{Somerville} R.~S., {Hopkins} P.~F., {Cox} T.~J., {Robertson} B.~E.,
  {Hernquist} L., 2008, \mnras, 391, 481

\bibitem[{{Somerville} \& {Kolatt}(1999)}]{Somerville1999}
{Somerville} R.~S., {Kolatt} T.~S., 1999, \mnras, 305, 1

\bibitem[{{Somerville} {et~al.}(2001){Somerville}, {Lemson}, {Sigad}, {Dekel},
  {Kauffmann}, \& {White}}]{Somerville2001}
{Somerville} R.~S., {Lemson} G., {Sigad} Y., {Dekel} A., {Kauffmann} G.,
  {White} S.~D.~M., 2001, \mnras, 320, 289

\bibitem[{{Springel}(2005)}]{Springel2005a}
{Springel} V., 2005, \mnras, 364, 1105

\bibitem[{{Sutherland} \& {Dopita}(1993)}]{Sutherland1993}
{Sutherland} R.~S., {Dopita} M.~A., 1993, \apjs, 88, 253

\bibitem[{{Ter Braak}(2006)}]{TerBraak2006}
{Ter Braak} C. J.~F., 2006, Stat. Comput., 16, 239

\bibitem[{{van den Bosch}(2002)}]{vandenBosch2002}
{van den Bosch} F.~C., 2002, \mnras, 331, 98

\bibitem[{{Warren} {et~al.}(1992){Warren}, {Quinn}, {Salmon}, \&
  {Zurek}}]{Warren1992}
{Warren} M.~S., {Quinn} P.~J., {Salmon} J.~K., {Zurek} W.~H., 1992, \apj, 399,
  405

\bibitem[Weinberg (2010a)]{BIE2010}
{Weinberg} M.~D., 2010a,  The UMass Bayesian Inference Engine, {\tt http://www.astro.umass.edu/BIE}

\bibitem[Weinberg (2010b)]{Weinberg2010}
{Weinberg} M.~D., 2010b, in preparation


\bibitem[{{Weinmann} {et~al.}(2006){Weinmann}, {van den Bosch}, {Yang}, \&
  {Mo}}]{Weinmann2006}
{Weinmann} S.~M., {van den Bosch} F.~C., {Yang} X., {Mo} H.~J., 2006, \mnras,
  366, 2

\bibitem[{{White} \& {Frenk}(1991)}]{White1991}
{White} S.~D.~M., {Frenk} C.~S., 1991, \apj, 379, 52

\bibitem[{{White} \& {Rees}(1978)}]{White1978}
{White} S.~D.~M., {Rees} M.~J., 1978, \mnras, 183, 341

\bibitem[{{Zhao} {et~al.}(2003{\natexlab{a}}){Zhao}, {Jing}, {Mo}, \&
  {B{\"o}rner}}]{Zhao2003a}
{Zhao} D.~H., {Jing} Y.~P., {Mo} H.~J., {B{\"o}rner} G., 2003{\natexlab{a}},
  \apjl, 597, L9

\bibitem[{{Zhao} {et~al.}(2009){Zhao}, {Jing}, {Mo}, \&
  {B{\"o}rner}}]{Zhao2009}
---, 2009, \apj, 707, 354

\bibitem[{{Zhao} {et~al.}(2003{\natexlab{b}}){Zhao}, {Mo}, {Jing}, \&
  {B{\"o}rner}}]{Zhao2003}
{Zhao} D.~H., {Mo} H.~J., {Jing} Y.~P., {B{\"o}rner} G., 2003{\natexlab{b}},
  \mnras, 339, 12

\end{thebibliography}

%%%%%%%%%%%%%%%%%%%%%%%%%%%%%%%%%%%%%%%%%%%%%%%%%%%%%%%%%%%%%%%%%%%%%%%%%%
\newpage
%%%%%%%%%%%%%%%%%%%%%%%%%%%%%%%%%%%%%%%%%%%%%%%%%%%%%%%%%%%%%%%%%%%%%%%%%%
\newpage
%%%%%%%%%%%%%%%%%%%%%%%%%%%%%%%%%%%%%%%%%%%%%%%%%%%%%%%%%%%%%%%%%%%%%%%%%%
\newpage
%%%%%%%%%%%%%%%%%%%%%%%%%%%%%%%%%%%%%%%%%%%%%%%%%%%%%%%%%%%%%%%%%%%%%%%%%%
\newpage
%%%%%%%%%%%%%%%%%%%%%%%%%%%%%%%%%%%%%%%%%%%%%%%%%%%%%%%%%%%%%%%%%%%%%%%%%%
\newpage

%%%%%%%%%%%%%%%%%%%%%%%%%%%%%%%%%%%%%%%%%%%%%%%%%%%%%%%%%%%%%%%%%%%%%%%%%%
\newpage
%%%%%%%%%%%%%%%%%%%%%%%%%%%%%%%%%%%%%%%%%%%%%%%%%%%%%%%%%%%%%%%%%%%%%%%%%
\newpage

%%%%%%%%%%%%%%%%%%%%%%%%%%%%%%%%%%%%%%%%%%%%%%%%%%%%%%%%%%%%%%%%%%%%%%%%%%
\newpage
%%%%%%%%%%%%%%%%%%%%%%%%%%%%%%%%%%%%%%%%%%%%%%%%%%%%%%%%%%%%%%%%%%%%%%%%%

\end{document}